\documentclass[lettersize,journal]{IEEEtran}
\IEEEoverridecommandlockouts
\usepackage{graphicx, amsmath, amsthm, amssymb, subcaption, url, cite, array, amsthm}
\usepackage[ruled, linesnumbered]{algorithm2e} 
\usepackage{algorithm2e, setspace}
\usepackage[font={small}]{caption}
\usepackage{hyperref}
\usepackage{xcolor}

\newtheorem{definition}{Definition}
\newtheorem{proposition}{Proposition}
\newtheorem*{remark}{Remark}

\begin{document}
\title{\huge Reconstructing Human Pose from Inertial Measurements:\\ A
Generative Model-based Compressive Sensing Approach}
\author{Nguyen Quang Hieu, Dinh Thai Hoang, Diep N. Nguyen, and Mohammad Abu Alsheikh

\thanks{This work was supported in part by the Australian Research Council under grants DE200100863 and DE210100651.
}
\thanks{N. Q. Hieu is with School of Information Technology and Systems, University of Canberra, Canberra, ACT 2617, Australia, and also with School of Electrical Data Engineering, University of Technology, Sydney, NSW 2007, Australia, emails: (hieu.nguyen-1@student.uts.edu.au; hieu.nguyen@canberra.edu.au)}
\thanks{D. T. Hoang and D. N. Nguyen are with School of Electrical and Data Engineering, University of Technology Sydney, NSW 2007, Australia, emails: 
(diep.nguyen@uts.edu.au; hoang.dinh@uts.edu.au).}
\thanks{M. A. Alsheikh is with Faculty of Science \& Technology, University of Canberra, Canberra, ACT 2617, Australia, email: mohammad.abualsheikh@canberra.edu.au.}
}
\maketitle
\begin{abstract}
The ability to sense, localize, and estimate the 3D position and orientation of the human body is critical in virtual reality (VR) and extended reality (XR) applications. This becomes more important and challenging with the deployment of VR/XR applications over the next generation of wireless systems such as 5G and beyond. In this paper, we propose a novel framework that can reconstruct the 3D human body pose of the user given sparse measurements from Inertial Measurement Unit (IMU) sensors over a noisy wireless environment.
Specifically, our framework enables reliable transmission of compressed IMU signals through noisy wireless channels and effective recovery of such signals at the receiver, e.g., an edge server. This task is very challenging due to the constraints of transmit power, recovery accuracy, and recovery latency. To address these challenges, we first develop a deep generative model at the receiver to recover the data from linear measurements of IMU signals. The linear measurements of the IMU signals are obtained by a linear projection with a measurement matrix based on the compressive sensing theory. The key to the success of our framework lies in the novel design of the measurement matrix at the transmitter, which can not only satisfy power constraints for the IMU devices but also obtain a highly accurate recovery for the IMU signals at the receiver. This can be achieved by extending the set-restricted eigenvalue condition of the measurement matrix and combining it with an upper bound for the power transmission constraint.
Our framework can achieve robust performance for recovering 3D human poses from noisy compressed IMU signals. Additionally, our pre-trained deep generative model achieves signal reconstruction accuracy comparable to an optimization-based approach, i.e., Lasso, but is an order of magnitude faster.
\end{abstract}


\section{Introduction}
\subsection{Motivation}
The ability to estimate human body movements plays a key role in emerging human-computer interaction paradigms such as virtual reality (VR) and extended reality (XR) \cite{liu2022recent}. By correctly estimating the 3D position and orientation of the human body, VR/XR applications such as gaming, virtual offices, and smart factories can offer a more interactive and immersive experience for users. Highly accurate solutions for estimating 3D human movements usually rely on images or videos, which typically require multi-camera calibrated systems \cite{liu2022recent, vicon2023}. However, the multi-camera systems are limited to capturing human outdoor activities (e.g., due to sensitive information conveyed in the images/videos) and severely degraded with poor lightning conditions \cite{liu2022recent}. Specifically, for VR/XR applications deployed over wireless systems, e.g., 5G and beyond, leveraging such images and videos from multi-camera systems for human body estimation purposes is costly in terms of bandwidth utilization and computing efficiency \cite{siriwardhana2021survey, sebire2023extended}. This demands a more effective approach to achieve highly accurate estimation of human body movements in VR/XR applications deployed over wireless systems \cite{behravan2022positioning}. 

Fortunately, the inertial measurement unit (IMU) (i.e., accelerometer, gyroscope, and magnetometer)  offers a promising solution to this problem. The systems based on IMU do not suffer from limitations in camera-based systems. The IMU sensors can track human movements by measuring the acceleration and orientation of human body parts, e.g., head orientation or arm/leg movement, regardless of image sensitivity information and lightning conditions, making them more suitable for indoor and outdoor VR/XR applications \cite{von2017sparse, xsens2023}. As the IMU sensors are typically worn on the body, e.g., wrists, head, or ankles, the information measured from the IMU can help to track the movement of the body segments relative to each other.
For example, utilizing IMU information such as the orientation of VR headsets can help the system better predict the user preferences in VR streaming applications \cite{hieu2023virtual, zhang2019drl360}. Moreover, acceleration readings from the IMU sensors can help to track user step count, thereby increasing the accuracy of outdoor pedestrian localization\cite{jiang2018ptrack}. Furthermore, combining IMU information with a kinematic model of the human body can simulate the entire body movement of the user in a complete positioning and sensing system \cite{xsens2023, guzov2021human}. With such enormous potential, the IMU sensors have been widely deployed as a standard setting inside mobile phones, tablets, VR headsets, and VR controllers. 

\subsection{Related Works}

Unlike solutions for reconstructing movements of independent parts of the human body, e.g., head or arm, estimating a full body movement of the user usually requires a set of IMU sensors placed on different parts of the body or attached to a suit \cite{xsens2023}. With a set of IMU sensors, ranging from 3 to 17 sensors, the full body movements can be fully reconstructed with the help of optimization-based techniques \cite{roetenberg2009xsens, von2017sparse} and learning-based techniques \cite{huang2018deep, yi2022physical, winkler2022questsim}. 
In \cite{roetenberg2009xsens}, a Kalman Filter was utilized to correct the kinematics of the 3D human model, given the joint uncertainties of sensor noise, angular velocity, and acceleration of the IMU sensors. 
In \cite{von2017sparse}, the authors proposed a new optimization approach based on exponential mapping, which transforms the orientation and acceleration values into equivalent energy functions. After carefully calibrating between the IMU sensors' coordinate frames and the 3D human body's coordinate frames, the optimization objective can be formulated as minimizing the set of energy functions over the entire sequence of collected data.

Different from offline optimization approaches \cite{roetenberg2009xsens, von2017sparse}, learning-based approaches can achieve real-time estimation based on pre-trained deep learning models \cite{huang2018deep, yi2022physical, winkler2022questsim}. In \cite{huang2018deep}, the authors proposed a deep learning approach based on a recurrent neural network that trains on the entire sequence of data at the training phase. During the testing phase, the pre-trained model can estimate the corrected body pose of the user in a shorter time window. In \cite{yi2022physical}, the authors extended this idea by using a recurrent neural network combined with a physics-aware motion optimizer that enhances the tracking accuracy for a longer time window. The authors in \cite{winkler2022questsim} reported similar advantages of using a physics-aware motion optimizer with fewer IMU sensors being used.

\subsection{Challenges and Proposed Solutions}

Although there has been significant effort in improving the precision of human pose estimation in 3D environments with IMU sensors\cite{roetenberg2009xsens, xsens2023, von2017sparse, huang2018deep, yi2022physical, winkler2022questsim}, there is a lack of human pose estimation approach for VR/XR applications deployed over wireless networks, where the estimation ability can be strictly constrained by channel quality, power transmission, and tradeoff between latency and accuracy of the solution.
Deploying the human pose estimation frameworks over the wireless networks is a non-trivial problem as the transmitted data is more exposed to channel noise, e.g., due to channel quality and channel interference. At this point, the existing works overlook the presence of noise in the received IMU data as the noise may significantly degrade the reconstruction accuracy of the human pose estimation task, resulting in a poor quality of experience for the user in a virtual environment. 
In addition, the current approaches do not consider the potential redundancy of the IMU data before transmitting it to the receiver. 
The potential redundancy of information from each IMU sensor, such as orientation and acceleration values at high frequencies (see Fig.~\ref{fig:fft-imu}), can further reduce the number of data samples that need to be transmitted. As a result, exploiting the data redundancy can enhance channel utilization and reduce power consumption for the wireless systems \cite{kong2016embracing}.

To the best of our knowledge, there is a lack of studies that address the above problems, i.e., the noisy IMU data transmission and the potential redundancy of the IMU data in reconstructing human pose over wireless systems.
To address these problems, we propose a novel framework based on compressive sensing and generative modeling.
On the one hand, the compressive sensing technique is utilized to down-sample the IMU signals before transmitting the signals over the wireless channel \cite{choi2017compressed}. Based on rigorous down-sampling (or projection) techniques, the compressive sensing framework is promising to reduce the redundant IMU signals.
As a result, this approach can not only reduce the energy consumption associated with IMU data acquisition and transmission but also enhance channel utilization as the transmitted data is in a more compressed form.
On the other hand, a generative model (e.g., a variational auto-encoder \cite{kingma2013auto}) deployed at the receiver can help the receiver make a robust estimation of the human body pose (e.g., through data denoising and data recovery capabilities), given the noisy compressed IMU measurements transmitted over a wireless channel. Unlike the optimization-based techniques \cite{roetenberg2009xsens, von2017sparse} and learning-based techniques \cite{huang2018deep, yi2022physical, winkler2022questsim}, our proposed generative model can handle noisy data more effectively and also can exploit the potential sparsity patterns in the data.
To summarize, the main contributions of our work are as follows:


\begin{itemize}
\item We propose an innovative framework based on compressive sensing and generative modeling techniques for human pose estimation from IMU sensors toward VR/XR applications deployed over wireless systems. The proposed framework can accurately recover the original IMU signals from noisy compressed signals transmitted over a wireless channel. The combination of compressive sensing and generative modeling generates potential benefits to the system such as enhancing channel utilization, effective data sensing, and power-efficient wireless communications. 
\item We develop a novel design for the measurement matrix at the transmitter, which helps to transform the high-dimensional signal into a compressed form. Our measurement matrix design extends the set-restricted eigenvalue condition of existing generative model-based compressive sensing approaches to a general setting, which considers the impacts of a wireless communication channel on data recovery. With rigorous analysis, we prove that our proposed measurement matrix enables our proposed framework to outperform other deep learning and optimization approaches, in terms of accuracy and latency of signal reconstruction process.
\item We show that the proposed framework can achieve signal reconstruction accuracy comparable to the optimization-based approach, i.e., Lasso, with an order of magnitude faster than Lasso. The fast reconstruction ability of the generative model makes it a promising solution for VR/XR applications with stringent latency requirements.
\item We demonstrate a practical use case of the generative model that can generate missing IMU signals, thus creating synthetic body movements for the users without using input IMU signals. This ability of the generative model is very useful for potential VR/XR applications over wireless systems as the missing input data usually happens due to the lossy nature of the wireless environment.
\end{itemize}

\begin{figure*}[t]
\centering
\includegraphics[width=1.0\linewidth]{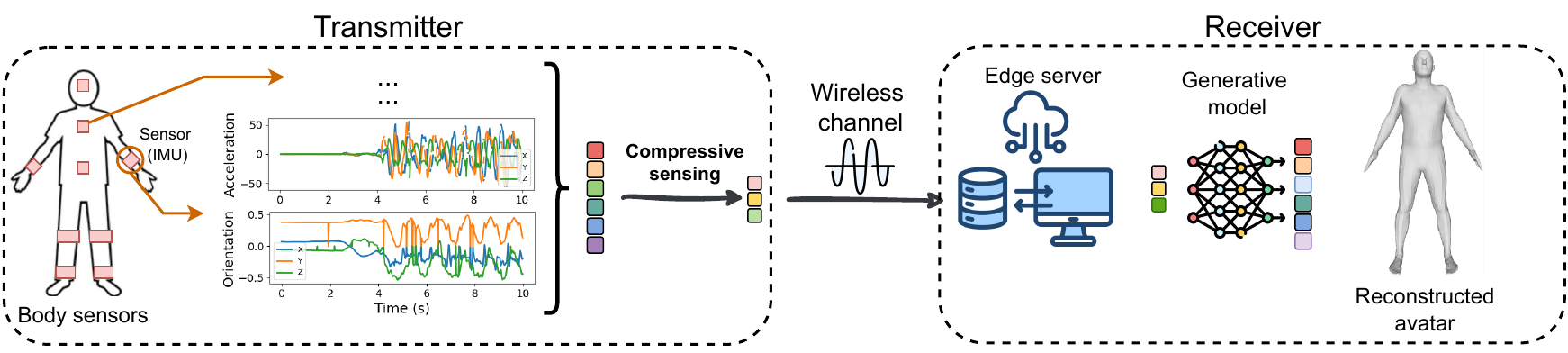}
\caption{An illustration of our proposed system model. A set of synchronized IMU sensors produces a sequence of data, e.g., orientation and acceleration, and compressive sensing down-samples the data sequence into a shorter sequence. The down-sampled sequence of IMU data is transmitted over a noisy channel. The receiver uses a deep generative model to recover the original data sequence from received signals.}
\label{fig:system-model}
\end{figure*}

The organization of the paper is as follows. Section \ref{sec:system-model} describes the overview of the system model and preliminaries of compressive sensing and generative modeling. In Section \ref{sec:problem-formulation}, we formulate the problem as a reconstruction error minimization problem, subject to a power transmission constraint. In Section \ref{sec:performance}, we extensively evaluate the performance of the proposed framework with other baselines, such as an optimization-based approach and a deep learning-based approach. We also show that the proposed generative model can generate missing IMU data features, thus directly creating smooth synthetic body movements for the users. Finally, Section \ref{sec:conclusion} concludes the paper.

\section{System Overview and Preliminaries}
\label{sec:system-model}

The proposed system model is illustrated in Fig.~\ref{fig:system-model}. At the transmitter side, the user's body is equipped with a set of 17 IMU sensors placed on standard positions as in commercial systems \cite{xsens2023}. The set of synchronized IMU sensors produces a sequence of data, e.g., orientation and acceleration, which is usually aggregated together at a central IMU node (e.g., IMU sensor placed on the user's spine). Compressive sensing down-samples the data sequence into a shorter sequence through matrix multiplication. The down-sampled data sequence is transmitted over a wireless noisy channel, e.g., a Gaussian channel. On the receiver side, the edge server uses a deep generative model to recover the original data sequence from the noisy down-sampled data sequence. From the recovered IMU data and a kinematic human body model (e.g., SMPL \cite{loper2023smpl}), the generative model can further generate the 3D avatar model with the corrected pose.

As described, the proposed framework consists of two main components that are (i) compressive sensing for the transmitter-receiver communication and (ii) a generative model for recovering the signals at the receiver, i.e., the edge server. In the following, we describe the fundamentals of compressive sensing, generative models, and generative model-based compressive sensing for an end-to-end learning system. 

\subsection{Compressive Sensing}
As illustrated in Fig.~\ref{fig:system-model}, we have a sequence of data is a real-valued, finite-length one-dimensional signal $\mathbf{x}^* \in \mathbb{R}^{n}$. With compressing sensing, we want to down-sample the signal $\mathbf{x^*}$ before transmitting it to the receiver. For that, we have a measurement matrix $\mathbf{A} \in \mathbb{R}^{m \times n}$ to make a linear projection from a higher dimensional vector $\mathbf{x}^* \in \mathbb{R}^n$ to a lower dimensional vector $\mathbf{y} \in \mathbb{R}^m$ ($m < n$). Usually, $n$ is referred to as the length of the original vector and $m$ is the number of measurements from that vector. 
In particular, the $m$-dimensional signal being transmitted over the channel is:
\begin{equation}
\mathbf{y} = \mathbf{Ax^*}.
\end{equation}

The received signal at the receiver can be corrupted by noise. In the case of a Gaussian channel, the received signal at the receiver is \cite{foucart2013invitation, choi2017compressed}:
\begin{equation}
\label{eq:hat-y}
\mathbf{\hat{y}} = \mathbf{Ax^*} + \boldsymbol{\eta},
\end{equation}
where $\boldsymbol{\eta} \in \mathbb{R}^m$ is a Gaussian noise vector with zero mean and $\sigma_N$ standard deviation, i.e., element $\eta_i$ ($i=1, 2, \ldots, m$) of $\boldsymbol{\eta}$ follows a Gaussian distribution $\eta_i \sim \mathcal{N}(0, \sigma_N^2)$. As  observed from equation (\ref{eq:hat-y}), the signal $\mathbf{\hat{y}} \in \mathbb{R}^m$ is a compressed form of $\mathbf{x}^* \in \mathbb{R}^n$.
To recover the signal $\mathbf{x}^*$ from the received signal $\mathbf{\hat{y}}$, the receiver needs to solve the following quadratically constrained optimization problem \cite{foucart2013invitation}:
\begin{subequations}
\label{eq:l1-min}
\begin{align}
\mathcal{P}_0: \quad & \min_{\mathbf{x}} \|\mathbf{x}\|_1, \\
\textrm{subject to} \quad & \|\mathbf{Ax} - \mathbf{\hat{y}}\|_2 \leq \|\boldsymbol{\eta}\|_2,
\end{align}
\end{subequations}
where the term $\|\mathbf{x}\|_p$ denotes the $l_p$ norm ($p= 0, 1, 2, \ldots$) of the vector $\mathbf{x}$, i.e., \cite{foucart2013invitation}
\begin{equation}
\|\mathbf{x}\|_p = (\sum_{j=1}^n |x_j|^p)^{1/p}.
\label{eq:lp-definition}
\end{equation}

The problem $\mathcal{P}_0$ in (\ref{eq:l1-min}) forms an underdetermined system, that is, a system in which there are multiple solutions to the system. To guarantee the unique recovery of the signal, compressive sensing relies on the two main assumptions about the signal $\mathbf{x}^*$ and the measurement matrix $\mathbf{A}$. First, the signal $\mathbf{x^*}$ has sparsity property. Second, the measurement matrix $\mathbf{A}$ satisfies specific conditions that are either Restricted Isometry Property (RIP) or Restricted Eigenvalue Condition (REC). The definitions of sparsity, RIP, and REC are as follows \cite{foucart2013invitation, choi2017compressed}.

\begin{figure}[t]
\centering
\includegraphics[width=0.9\linewidth]{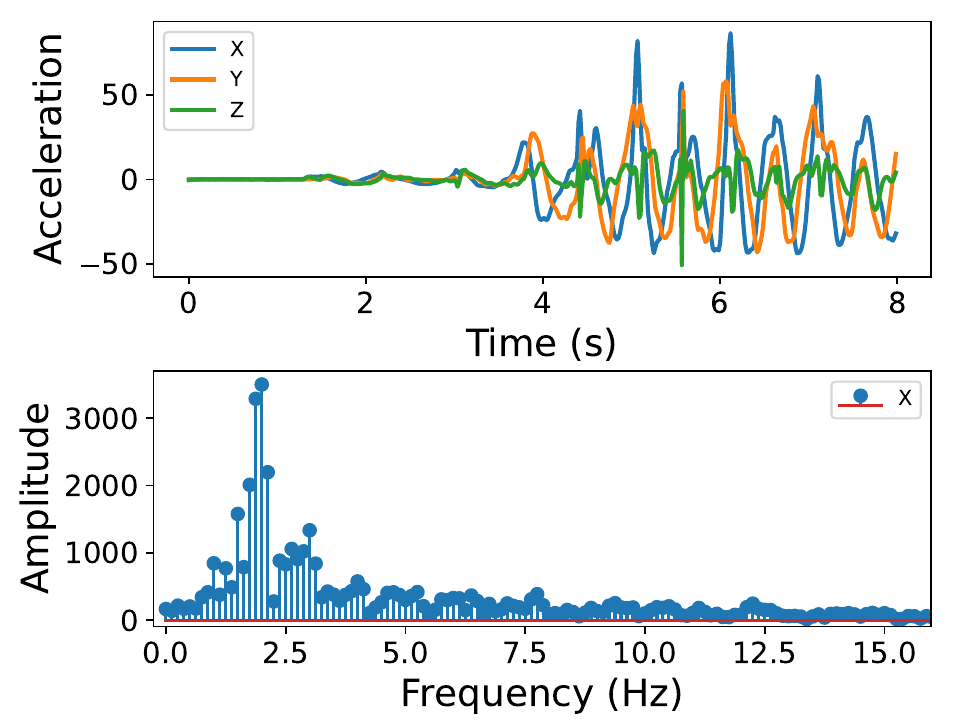}
\caption{Illustration of acceleration reading from an IMU sensor placed on the left wrist of the user (top figure) and the Fast Fourier Transform (FFT) of the x-axis acceleration data (bottom figure). The FFT reveals nearly $k$-sparse property of the IMU signal in which a few low-frequency coefficients have dominant values. As a result, the redundancy of the data can be approximated by considering the $k$ largest coefficients and assuming the rest coefficients are zero.}
\label{fig:fft-imu}
\end{figure}

\begin{definition}[Sparsity]
\label{def:sparsity}
The support of a vector $\mathbf{x} \in \mathbb{R}^n$ is the index set of its nonzero entries, i.e.,
\begin{equation*}
\sup(\mathbf{x}) := \big\{j \in \{1, 2, \ldots, n\}: x_j \neq 0 \big\}.
\end{equation*}
The vector $\mathbf{x} \in \mathbb{R}^n$ is called $k$-sparse if at most $k$ of its entries are nonzero, i.e., if
\begin{equation*}
\|\mathbf{x}\|_0 = \mathbf{card}\big(\sup (\mathbf{x})\big) \leq k,
\end{equation*}
where $\mathbf{card}(\cdot)$ is the cardinality (number of elements) and $\|\cdot\|_0$ is $l_0$ norm.
\end{definition}
In practice, the sparsity property of the interested signal $\mathbf{x}$ is usually relaxed to nearly $k$-sparse, meaning that there are $n-k$ entries of the vector $\mathbf{x}$ are approximately zero.  In Fig. \ref{fig:fft-imu}, we illustrate the nearly $k$-sparse acceleration data from the IMU dataset in \cite{huang2018deep}. 
Note that we use the Fast Fourier Transform (FFT) in Fig.~\ref{fig:fft-imu} for illustration purposes only, and our proposed learning algorithm will not utilize the FFT.

\begin{definition}[Restricted Isometry Property]
\label{def:rip}
Let $S_k\subset \mathbb{R}^n$ be the set of $k$-sparse vectors. For some parameter $\delta \in (0, 1)$, a matrix $\mathbf{A} \in \mathbb{R}^{m \times n}$  is said to satisfy $\text{RIP}(k, \delta)$ if $ \forall \mathbf{x} \in S_k$,
\begin{equation*}
 (1 - \delta) \|\mathbf{x}\|_2 \leq \|\mathbf{Ax}\|_2 \leq (1 + \delta)\|\mathbf{x}\|_2.
\end{equation*}
\end{definition}

\begin{definition}[Restricted Eigenvalue Condition]
\label{def:rec}
Let $S_k\subset \mathbb{R}^n$ be the set of $k$-sparse vectors. For some parameter $\gamma > 0$, a matrix $\mathbf{A} \in \mathbb{R}^{m \times n}$  is said to satisfy $\text{REC}(k, \gamma)$ if $ \forall \mathbf{x} \in S_k$,
\begin{equation*}
\|\mathbf{Ax}\|_2 \geq \gamma \|\mathbf{x}\|_2.
\end{equation*}
\end{definition}

Intuitively, RIP implies that $\mathbf{A}$ approximately preserves Euclidean norms ($l_2$ norms) for sparse vectors, and REC implies that sparse vectors are far from the nullspace of $\mathbf{A}$ \cite{dhar2018modeling}. Given the sparsity of $\mathbf{x}^*$ and RIP/REC property of the chosen matrix $\mathbf{A}$, it has been shown that the recovered signal $\mathbf{\hat{x}}$ is the unique solution of the problem $\mathcal{P}_0$ in (\ref{eq:l1-min}), i.e., $\mathbf{\hat{x}} \approx \mathbf{x}^*$ \cite{foucart2013invitation}. As the sparsity of the signal depends on the natural domain of the signal, the solution for $\mathcal{P}_0$ depends on two aspects that are (i) the choice of measurement matrix $\mathbf{A}$ and (ii) the choice of recovery method, i.e., optimization solver for $\mathcal{P}_0$. In conventional compressive sensing methods, the common choices for such aspects are (i) a Gaussian matrix and (ii) a convex optimization solver like Lasso (Lease absolute shrinkage and selection operator) \cite{tibshirani1996regression, choi2017compressed}. Note that in this work, we do not explicitly analyze the sparsity of the IMU signal but rely on approximation methods, such as Lasso and generative models, to solve the optimization problem. In this way, we do not need to pay the extra cost of signal processing through transformations, such as Fourier transform or Wavelet transform, at the transmitter \cite{choi2017compressed, foucart2013invitation, bora2017compressed, dhar2018modeling}.

By using the definition of the $l_p$ norm in (\ref{eq:lp-definition}), $\|\mathbf{x}\|_1$ is a convex function, and $\mathcal{P}_0$ is a $l_1$ minimization problem with quadratic constraint.
The solution of $\mathcal{P}_0$ is equivalent to the output of the Lasso, which consists of solving $\mathcal{P}_1$, for some parameter $\tau \geq 0$ \cite{foucart2013invitation}:
\begin{subequations}
\label{eq:lasso-min}
\begin{align}
\mathcal{P}_1: \quad & \min_{\mathbf{x}} \|\mathbf{Ax} - \mathbf{\hat{y}}\|_2, \\
\textrm{subject to} \quad & \|\mathbf{x}\|_1 \leq \tau.
\end{align}
\end{subequations}
In practice, the solution of Lasso is equivalent to solving the Lagrangian of the problem $\mathcal{P}_1$ above, for some parameter $\lambda \geq 0$, i.e., \cite[Equation 3.1]{foucart2013invitation}:
\begin{equation}
\label{eq:lasso-min-lambda}
\min_{\mathbf{x}}\|\mathbf{Ax} - \mathbf{\hat{y}}\|_2^2 + \lambda \|\mathbf{x}\|_1.
\end{equation}

Intuitively, the $l_1$ penalty term $\lambda \|\mathbf{x}\|_1$ in (\ref{eq:lasso-min-lambda}) enforces sparsity (Definition \ref{def:sparsity}) by adding penalty proportional to the absolute values of the coefficients of $\mathbf{x}$. As a result, the sparsity assumption ($k$-sparse or nearly $k$-sparse) in the structure of the signal has an impact on the performance of the Lasso solver. Recall that in this work, we do not explicitly analyze the sparsity of the IMU signal, which usually causes extra costs through the Fourier/Wavelet transform at the transmitter. In addition, the solutions relying on the sparsity assumption are known to yield poor recovery performance when the linear measurement is not sufficient (i.e., the value of $m$ is too small), or the considered signal has a small number of dimensions (i.e., the value of $n$ is not sufficiently large) \cite{bora2017compressed, dhar2018modeling}. This motivates us to utilize generative models, such as variational auto-encoders (VAEs) \cite{kingma2013auto} and generative adversarial networks (GANs) \cite{goodfellow2014generative}, as an alternative for the use of sparsity assumption with a convex optimization solver like Lasso. 

\subsection{Generative Models}
Generative models are a type of machine learning that can be used for modeling the complex distribution of large-scale datasets. In the context of compressive sensing, a generative model can be used to estimate the distribution of the input signals. After the generative model is trained with a training set, it can generate a new data sample that is similar to the samples drawn from the original set \cite{kingma2019introduction}. Intuitively, the generative model can learn and synthesize the underlying distribution of the high dimensional and complex data, which eliminates the sparsity assumption about the data structure of conventional compressive sensing techniques. 

A generative model describes a probability density function $p$: $\mathcal{X} \rightarrow \mathbb{R}$ ($\mathcal{X}$ is a finite set) through an unobserved, or ``latent", variable $\mathbf{z}$. The probability density function is then calculated by:
\begin{equation}
p(\mathbf{x}) = \int_{\mathbf{z}} p(\mathbf{x|z}) p(\mathbf{z}) d \mathbf{z},
\label{eq:pdf-z}
\end{equation}  
where $\forall \mathbf{x} \in \mathcal{X}$, the probability $p(\mathbf{z})$ is the prior, and the forward probability $p(\mathbf{x|z})$ is the likelihood \cite{kingma2019introduction}. 
In practice, this probability density function is usually parameterized by a model $\boldsymbol{\theta}$ (e.g., a deep neural network). 
In such a case, equation (\ref{eq:pdf-z}) can be rewritten as follows \cite{kingma2019introduction}:
\begin{equation}
p_{\boldsymbol{\theta}}(\mathbf{x}) = \int_{\mathbf{z}} p_{\boldsymbol{\theta}}(\mathbf{x|z}) p(\mathbf{z}) d\mathbf{z}.
\label{eq:pdf-theta-z}
\end{equation}
The integral in (\ref{eq:pdf-theta-z}) cannot easily be computed as the likelihood $p(\mathbf{x|z})$ is computationally expensive with conventional methods such as maximum likelihood, especially for large-scale datasets. In this work, we develop our generative model based on a popular class of generative models, which are called variational auto-encoders (VAEs), first introduced in \cite{kingma2013auto}. As opposed to other generative models such as GANs \cite{goodfellow2014generative}, VAEs can generate more dispersed samples over the data and can learn complex data distributions \cite{kingma2019introduction}. In addition, VAEs are better for data inference, which is suitable for our generative model that wants to exploit the hidden ``sparsity" patterns in the IMU data. 

In VAEs, besides the likelihood parameterized by a decoder (deep neural network), the probability density function $p_{\boldsymbol{\theta}}(\mathbf{x})$ is conditioned through an encoder parameterized by another deep neural network $q_{\boldsymbol{\phi}}(\mathbf{z|x})$. The encoder approximates the true but intractable posterior $p_{\boldsymbol{\theta}}(\mathbf{z|x})$. To train a VAE, we optimize a variational lower bound on $\log p_{\boldsymbol{\theta}}(\mathbf{x})$, called evidence lower-bound (ELBO). It is defined as follows \cite{kingma2019introduction}:
\begin{equation}
\mathcal{L}_{\boldsymbol{\theta}, \boldsymbol{\phi}}(\mathbf{x}) = \int_{\mathbf{z}} q_{\boldsymbol{\phi}}(\mathbf{z|x}) \log \frac{p_{\boldsymbol{\theta}}(\mathbf{x|z}) p(\mathbf{z})}{q_{\boldsymbol{\phi}}(\mathbf{z|x)}} d \mathbf{z}.
\label{eq:elbo}
\end{equation}

The newly introduced density function $q_{\boldsymbol{\phi}}(\mathbf{z|x})$ is referred to as the variational (approximate) posterior with $\boldsymbol{\phi}$ defined as the variational parameters. 

As the encoder $q_{\boldsymbol{\phi}}(\mathbf{z|x})$ is used to approximate the posterior $p_{\boldsymbol{\theta}}(\mathbf{z|x})$, exact sampling from the posterior is straightforward through an unbiased Monte Carlo estimate of $\mathcal{L}$ \cite{kingma2019introduction}:
\begin{equation}
\mathcal{\hat{L}}_{\boldsymbol{\theta, \phi}}(\mathbf{x}) = \log \frac{p_{\boldsymbol{\theta}}(\mathbf{x|z}) p(\mathbf{z})}{q_{\boldsymbol{\phi}}(\mathbf{z|x)}}, \text{ where } \mathbf{z} \leftarrow q_{\boldsymbol{\phi}}(\cdot|\mathbf{x}).
\label{eq:mote-carlo-estimation}
\end{equation}
The notation $\mathbf{z} \leftarrow q_{\boldsymbol{\phi}}(\cdot | \mathbf{x})$ means that $\mathbf{z}$ is sampled from the approximate posterior distribution $q_{\boldsymbol{\phi}}(\mathbf{z|x})$. If the process used to generate $\mathbf{z}$ from $q_{\boldsymbol{\phi}}(\mathbf{z|x})$ is differentiable with respect to $\boldsymbol{\phi}$, the function $\mathcal{\hat{L}}$ can be differentiated with respect to $\boldsymbol{\theta}$ and $\boldsymbol{\phi}$ by using a stochastic gradient decent estimator. Once $\mathcal{L}$ in (\ref{eq:elbo}) is optimized, we can approximate the true probability density function $p(\mathbf{x})$ through the learned neural network with parameters $\boldsymbol{\theta}$, i.e., $p_{\boldsymbol{\theta}}(\mathbf{x}) \approx p(\mathbf{x})$. In other words, we can generate the new data samples from the learned probability density function.

\begin{figure*}[t]
\centering
\includegraphics[width=0.9\linewidth]{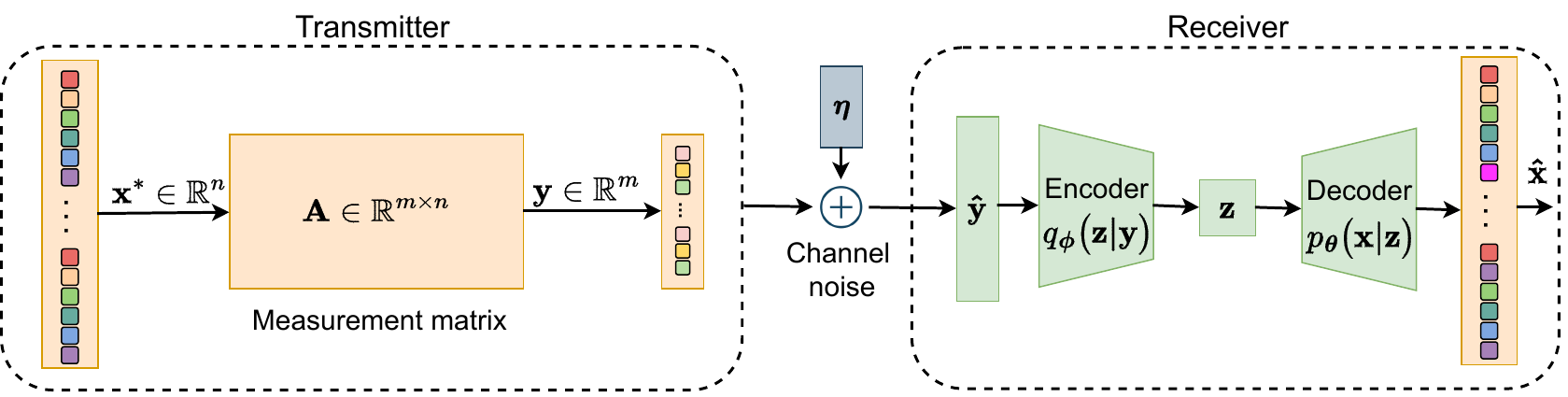}
\caption{The proposed CS-VAE learning algorithm with a novel measurement matrix at the transmitter and the generative model, i.e., a VAE, at the receiver. The transmitted signal at the transmitter is the $m$-dimensional vector $\mathbf{y}$, which is a compressed version of the original $n$-dimensional vector $\mathbf{x}^*$. At the receiver, the VAE recovers the original signal, i.e., $\mathbf{\hat{x}} \approx \mathbf{x}^*$, from a noisy and compressed measurement $\mathbf{\hat{y}}$.}
\label{fig:vae-training}
\end{figure*}

\subsection{Generative Model-based Compressive Sensing}
In the context of compressive sensing, the data sample $\mathbf{x}$ from the training set is, however, not fully observable, i.e., our model can only observe the noisy compressed or down-sampled version $\mathbf{\hat{y}}$. Replacing $\mathbf{x}$ with $\mathbf{y} = \mathbf{Ax}$, the unbiased Monte Carlo estimation of the ELBO in (\ref{eq:mote-carlo-estimation}) is rewritten as:
\begin{equation}
\mathcal{\hat{L}}_{\boldsymbol{\theta, \phi}}(\mathbf{y}) = \log \frac{p_{\boldsymbol{\theta}}(\mathbf{Ax|z}) p(\mathbf{z})}{q_{\boldsymbol{\phi}}(\mathbf{z|Ax})}, \text{ where } \mathbf{z} \leftarrow q_{\boldsymbol{\phi}}(\cdot|\mathbf{y} = \mathbf{\hat{y}}).
\label{eq:monte-carlo-y}
\end{equation}

As observed from the above equation, the generative model cannot directly generate the data sample $\mathbf{x}$ from the compressed observation $\mathbf{\hat{y}}$ without prior knowledge of the measurement matrix $\mathbf{A}$. In other words, the measurement matrix $\mathbf{A}$ is assumed to be known at the generative model \cite{bora2017compressed, dhar2018modeling}. Recall that the generative model is deployed at the receiver, therefore, the assumption about sharing prior information, e.g., a codebook, between the transmitter and receiver, is commonly used in the source and channel coding methods \cite{cover2006elements, feizi2010compressive}. Given the setting above, the solution of (\ref{eq:lasso-min}) is equivalent to the output of the generator $G(\mathbf{z})$ of the problem $\mathcal{P}_2$, for some $\upsilon \geq 0$, as follows \cite{dhar2018modeling}:
\begin{subequations}
\begin{align}
\mathcal{P}_2: \quad & \min_{\mathbf{z}} \|\mathbf{A}G(\mathbf{z}) - \mathbf{\hat{y}}\|_2, \\
\textrm{subject to} \quad & \|G(\mathbf{z})\|_1 \leq \upsilon.
\end{align}
\label{eq:generative-cs-min}
\end{subequations}

The generator $G(\mathbf{z})$ is defined as a function $G:\mathbb{R}^k \rightarrow \mathbb{R}^n$ mapping a latent vector $\mathbf{z}$ to the mean of the conditional distribution $p_{\boldsymbol{\theta}}(\mathbf{x|z})$. Given the observation $\mathbf{\hat{y}}$ as the input of the model, the latent vector $\mathbf{z}$ is obtained by sampling from the posterior distribution $q_{\boldsymbol{\phi}}(\cdot | \mathbf{y})$ in (\ref{eq:monte-carlo-y}). After that, the generator $G(\mathbf{z})$ produces the output vector $\mathbf{\hat{x}}$ from this latent vector $\mathbf{z}$, i.e., $G(\mathbf{z}) = \mathbf{\hat{x}}$. As a result, the minimizer $\mathbf{z}^*$ of the optimization problem of $\mathcal{P}_2$ in (\ref{eq:generative-cs-min}) makes $G(\mathbf{z}^*) \approx \mathbf{x}^*$ \cite{bora2017compressed}.

In comparison with (\ref{eq:lasso-min}), the variable vector $\mathbf{x}$ is now replaced by the generative function $G(\mathbf{z})$ in (\ref{eq:generative-cs-min}). As a result, when one can optimize the objective $\mathcal{P}_2$ in (\ref{eq:generative-cs-min}), the generator $G(\mathbf{z})$ can generate the new samples which are similar to the original vector $\mathbf{x}^*$. Recall that under the compressive sensing setting, our generative model does not observe the full observation of the signals as in the conventional setting of generative modeling, i.e., learning directly $p_{\boldsymbol{\theta}}(\mathbf{x}) \approx p(\mathbf{x})$. In the compressive sensing setting, the generative model can only observe the noisy and compressed signal $\mathbf{\hat{y}}$. Therefore, the optimization objective defined in (\ref{eq:generative-cs-min}) is to indirectly optimize the generative model via the observation $\mathbf{\hat{y}}$, given the measurement matrix $\mathbf{A}$. Thereafter, the space of signals that can be recovered with the generative model is given by the range of the generator function, i.e.,

\begin{equation}
S_G = \{G(\mathbf{z}): \mathbf{z} \in \mathbb{R}^k\}.
\label{eq:set-G}
\end{equation}


As the range of the signals is now transformed into the latent space $\mathbf{z} \in \mathbb{R}^k$, the RIP and REC properties of the measurement matrix $\mathbf{A}$ no longer guarantee the accuracy of the recovered signals. With the generative model-based compressive sensing, the measurement matrix $\mathbf{A}$ is required to satisfy a Set-Restricted Eigenvalue Condition (S-REC), which is a generalized version of REC \cite{bora2017compressed}, i.e.,

\begin{definition}[Set-Restricted Eigenvalue Condition]
Let $S \subseteq \mathbb{R}^n$, for some parameters $\gamma > 0$ and  $\kappa \geq 0$, a matrix $\mathbf{A} \in \mathbb{R}^{m \times n}$ is said to satisfy the $\text{S-REC}(S, \gamma, \kappa)$ if $\forall \mathbf{x}_1, \mathbf{x}_2 \in S$,
\begin{equation*}
\|\mathbf{A(x_1 - x_2)}\|_2 \geq \gamma \|\mathbf{x_1 - x_2}\|_2 - \kappa.
\end{equation*}
\label{def:s-rec}
\end{definition}
Intuitively, the S-REC property generalizes the REC property to an arbitrary set of vectors $S$ instead of considering the set of approximately sparse vectors $S_k$ \cite{dhar2018modeling}. This generalization makes S-REC a nice property for solving a compressive sensing problem with a stochastic gradient estimator via deep neural networks.

\section{Problem Formulation and Proposed Learning Algorithm}
\label{sec:problem-formulation}

In this section, we utilize the generative model-based compressive sensing framework for our system model in Fig.~\ref{fig:system-model}. The presence of the communication channel between the transmitter and the receiver makes the reconstruction of IMU signals with generative model-based compressive sensing much more challenging. In particular, using the measurement matrices in \cite{bora2017compressed} and \cite{dhar2018modeling} cannot guarantee the power constraint of the transmitter. Conventional normalization techniques like $l_2$ normalization \cite{bourtsoulatze2019deep} are not applicable as they yield nonlinear projection from $\mathbf{x}$ to $\mathbf{y}$, thus making the receiver cannot recover the original signal. For this, we propose a new measurement matrix that (i) ensures the power constraint for the transmitter and (ii) satisfies the S-REC property of generative model-based compressive sensing. The learning algorithm with the newly designed measurement matrix is described in the following.

The proposed learning process, which we refer to as ``CS-VAE" (Compressive Sensing-based Variational Auto-Encoder), is illustrated in Fig.~\ref{fig:vae-training}. At the transmitter, we have the vector $\mathbf{x}^* \in \mathbb{R}^n$ and a measurement matrix $\mathbf{A} \in \mathbb{R}^{m \times n}$. The output of the measurement matrix is the signal $\mathbf{y} \in \mathbb{R}^m$ in which $m < n$. The signal $\mathbf{y = Ax^*}$ is subjected to the power constraint at the transmitter, i.e., $\frac{1}{m}  \|\mathbf{y}\|_2^2 \leq P_T$, where $P_T$ is the transmission power constraint on a single channel use \cite{feizi2010compressive, saidutta2021joint}.  Details of the power constraint for our framework are further discussed in Appendix \ref{appendix}.
In particular, the optimization problem of the proposed learning model is similar to (\ref{eq:generative-cs-min}) with an additional power constraint as follows:

\begin{subequations}
\begin{align}
\mathcal{P}_3: \quad &  \min_{\mathbf{z}} \|\mathbf{A}G(\mathbf{z}) - \mathbf{\hat{y}}\|_2, \\
\textrm{subject to} \quad & \frac{1}{m} \|\mathbf{y}\|_2^2 \leq P_T, \\
\quad & \|G(\mathbf{z})\|_1 \leq \upsilon.
\end{align}
\label{eq:main-optimization}
\end{subequations}

The power constraint in (\ref{eq:main-optimization}b) poses additional challenges in designing the measurement matrix $\mathbf{A}$ to ensure that the recovered signal is unique and similar to the original signal. Specifically, this is a very challenging quadratically constrained problem \cite{foucart2013invitation}, and designing a measurement matrix that satisfies the duo-constraint, i.e., S-REC property and power constraint, has not been investigated in the literature. Existing generative model-based compressive sensing approaches \cite{bora2017compressed, dhar2018modeling, jalal2021robust, mardani2018deep} cannot be directly applied to this problem. To address this duo-constraint optimization problem $\mathcal{P}_3$ in (\ref{eq:main-optimization}), we first design a new measurement matrix $\mathbf{A}$ in Proposition \ref{prop:main} that makes $\mathbf{y = Ax}$ satisfy the power constraint $\frac{1}{m}\|\mathbf{y}\|_2^2 \leq P_T$. After the power constraint is eliminated, we use the Lagrangian of $\mathcal{P}_3$ as a loss function to train the generative model in a similar manner as in  \cite{bora2017compressed}.

The proposed measurement matrix for the problem $\mathcal{P}_3$ is stated as follows.
\begin{proposition}[S-REC with power constraint]
The recovered signal obtained by the generative model-based compressive sensing method under the power constraint is guaranteed to be a unique solution if 
\begin{itemize}
\item $\mathbf{A}$ satisfies S-REC property, and
\item Each element $A_{ij}$ (element $j$-th of the $i$-th row) of $\mathbf{A}$ is drawn i.i.d from a Gaussian distribution with zero mean and variance $\sigma_a^2 = \frac{P_T}{n^2 d^2 (d \sigma_x + \mu_x)^2}$, i.e.,
\end{itemize}
\begin{equation*}
A_{ij} \sim \mathcal{N}\Big(0,\frac{P_T}{n^2 d^2 (d \sigma_x + \mu_x)^2}\Big),
\end{equation*}
where $\sigma_x^2$ and $\mu_x$ are the statistical variance and mean of the source signals $\mathbf{x} \in \mathbb{R}^n$, respectively, and $d > 0$ is a real number derived from the Chebyshev's inequality.
\label{prop:main}
\end{proposition}

The proof of Proposition \ref{prop:main} can be found in Appendix \ref{appendix}.

\begin{remark}
The normal distribution used to generate the random matrix $\mathbf{A}$ in Proposition \ref{prop:main} contains the mean and variance values of the source signals $\mathbf{x}$.  This assumption of knowing statistical variance and mean values of the source signals is common in source-channel coding schemes \cite{cover2006elements}. For example, in the case of i.i.d Gaussian source with power constraint $P$, these values are $\sigma_x^2 \approx P$ and $\mu_x = 0 $ \cite[Section 9.1]{cover2006elements}. In the view of compressive sensing as a source-channel coding scheme, $\mathbf{A}$ can be considered as an encoding function \cite{feizi2010compressive}. In addition, as training deep learning models usually requires access to the training set for pre-processing and learning, the assumption of knowing the mean and variance values of the signals is more reasonable and practical than using the i.i.d Gaussian source.
Another parameter in Proposition \ref{prop:main} is $d > 0$, a real number that restricts the random variable $x_j$ (element $j$-th of the source signal $\mathbf{x}$) to the interval $[\mu_x - d \sigma_x, \mu_x + d \sigma_x]$.
\textcolor{black}{
Given the measurement matrix $\mathbf{A}$ in Proposition \ref{prop:main}, the recovered signal of the optimization problem $\mathcal{P}_3$ is guaranteed to be the unique solution if (i) $\mathbf{A}$ satisfies the power constraint with probability greater than $1 - \frac{1}{d^2}$ and (ii) $\mathbf{A}$ satisfies the S-REC property with probability greater than $1 - 2 \exp \Big(\frac{(-2 \gamma m \sqrt{n} - m \kappa)^2}{4c \|\mathbf{A}\|_2^2}\Big)$. 
}
Details of the parameters are discussed in Appendix \ref{appendix}.
\end{remark}

As the measurement matrix $\mathbf{A}$ is designed based on the Proposition \ref{prop:main}, the power constraint in (\ref{eq:main-optimization}) is guaranteed and thus can be reduced. As a result, the optimization problem in (\ref{eq:main-optimization}) is equivalent to solving the following problem:
\begin{equation}
 \mathcal{P}_4: \min_{\mathbf{z}} \|\mathbf{A} G(\mathbf{z}) - \mathbf{\hat{y}}\|_2^2 + \lambda \|G(\mathbf{z})\|_1,
\label{eq:main-optimization-lambda}
\end{equation}
where $\lambda$ is the Lagrange multiplier.
As $\mathbf{z}$ is differentiable with respect to the generative model's parameters (e.g., using reparameterization trick \cite{kingma2019introduction}), one can use the loss function based on (\ref{eq:main-optimization-lambda}) to train the generative model \cite{dhar2018modeling}. 
Once the generative model is trained to obtain the solution for (\ref{eq:main-optimization}), denoted by $\mathbf{z}^*$, the reconstruction error can be bounded with probability $1 - e^{-\Omega(m)}$ by \cite{bora2017compressed}:
\begin{equation}
\label{eq:bounded-gz}
\|G(\mathbf{z}^*) - \mathbf{x}^*\|_2 \leq 6 \min_{\mathbf{z} \in \mathbb{R}^k} \|G(\mathbf{z}) - \mathbf{x}^*\|_2 + 3 \|\boldsymbol{\eta}\|_2 + 2 \epsilon,
\end{equation}
where $\epsilon$ is an additive error term caused by the use of gradient decent-based optimizers. 

Pseudo codes of our training algorithm are in Algorithm \ref{algo:cs-vae}, which can be described as follows. We first initialize the measurement matrix $\mathbf{A}$ following the Proposition \ref{prop:main}, together with random parameters for the inference network, i.e., encoder, $q_{\boldsymbol{\phi}}(\mathbf{z|y})$, and the generative model, i.e., decoder, $p_{\boldsymbol{\theta}}(\mathbf{x|z})$ (i.e., lines 1-3 of the Algorithm \ref{algo:cs-vae}). For each training loop, a batch of sample $\mathbf{x}^*$ is i.i.d sampled from the training set (line 5). The input of the VAE's encoder is $\mathbf{\hat{y}}$ obtained by using (\ref{eq:hat-y}) (line 6). The latent vector $\mathbf{z}$ is obtained in line 7, and the output of the VAE's decoder is $\mathbf{\hat{x}}$ in line 8. After that, the training loss $L(\mathbf{z})$ of the VAE can be computed as in line 9, which optimizes the problem $\mathcal{P}_4$ using the Adam optimization solver. After training, the pre-trained VAE can reconstruct signal $\mathbf{\hat{x}} = G(\mathbf{z}^*)$ ($\mathbf{z}^*$ is fixed during testing with the test set) with reconstruction error bounded by (\ref{eq:bounded-gz}) (line 14).

\begin{algorithm}[t]
\caption{CS-VAE: Training VAE to reconstruct signals from noisy compressed measurements}
\label{algo:cs-vae}
 \textbf{Input}: 
 Initialize measurement matrix $\mathbf{A}$ that satisfies Proposition \ref{prop:main}. \\
 Initialize encoder of the VAE with inference model $q_{\boldsymbol{\phi}}(\mathbf{z|y})$. \\
 Initialize decoder of the VAE with generative model $p_{\boldsymbol{\theta}}(\mathbf{x|z})$. \\
 \For{\text{t = 0, 1, 2}, $\ldots$}{
  Sample $\mathbf{x}^*$ from the training set. \\
  Obtain $\mathbf{\hat{y}}$ from (\ref{eq:hat-y}). \\
  Obtain latent vector $\mathbf{z} \leftarrow q_{\boldsymbol{\phi}}(\cdot | \mathbf{y})$, with $\mathbf{y} = \mathbf{\hat{y}}$. \\
  Obtain $\mathbf{\hat{x}} \leftarrow p_{\boldsymbol{\theta}}(\mathbf{x|z})$ at the output of the generator $G(\mathbf{z})$. \\
  Compute the loss based on (\ref{eq:main-optimization-lambda}), i.e., 
  \begin{equation}
  L(\mathbf{z}) = \|\mathbf{A} G(\mathbf{z}) - \mathbf{\hat{y}} \|_2^2 + \lambda \|G(\mathbf{z})\|_1,
  \end{equation} \\
  Update the neural network parameters by using backpropagation with the loss $L(\mathbf{z})$.
 }
  \textbf{Output}: $\mathbf{z} \rightarrow \mathbf{z}^*$. \\
  Reconstructed signal $\mathbf{\hat{x}} = G(\mathbf{z}^*)$. \\
  Reconstruction error: $\|G(\mathbf{z}^*) - \mathbf{x}^*\|_2$ bounded by (\ref{eq:bounded-gz}).
\end{algorithm}
 
\section{Performance Evaluation}
\label{sec:performance}
\subsection{Dataset and Simulation Settings}
\subsubsection{Dataset and VAE's Parameters}

We use the IMU data from the DIP-IMU dataset \cite{huang2018deep}, designed specifically for capturing 3D body human motion with calibrated IMU sensors. The dataset contains acceleration and orientation information of 17 IMU sensors placed on the participants. The entire dataset consists of 64 data sequences of 10 participants, equivalent to 330,178 frames of motion under various activities. The frames of motion are recorded at the rate of 60 frames per second. 
\textcolor{black}{
The activities can be divided into different categories that are upper body (e.g., arm raises, stretches, swings, and arm crossing), lower body (e.g., leg raises, squats, and lunges), locomotion (e.g., walking, sidesteps, and crossing legs), freestyle (e.g., jumping jacks, kicking, and boxing), and interaction (e.g., interacting with everyday objects such as keyboard, mobile phones, and grabbing objects). More details of action categories and their corresponding number of data frames can be found in Table 6 of reference \cite{huang2018deep}.
}

The output of the $j$-th IMU is a combination of orientation information, denoted by $\mathbf{o}^{(j)}_t \in \mathbb{R}^{9}$, and acceleration information, denoted by $\mathbf{a}^{(j)}_t \in \mathbb{R}^{3}$. One frame in the dataset at time step $t$ is denoted by $\mathbf{x}_t = \big[\mathbf{o}^{(1)}_t, \mathbf{a}^{(1)}_t, \mathbf{o}^{(2)}_t, \mathbf{a}^{(2)}_t, \ldots, \mathbf{o}^{(17)}_t, \mathbf{a}^{(17)}_t\big]$. As a result, one frame $\mathbf{x}_t$ in the dataset has $17 \times 9 + 17 \times 3 = 204$ features. We use the data sequences collected from 8 participants as the training set, and the data sequences collected from the other 2 participants as the test set. After removing all the data samples that have missing features (i.e., NaN values), the training set and test set contain 220,076 and 56,990 data samples, respectively. To stabilize the training process, we normalize the data $\mathbf{x}_t$ in the training and test sets within the range $(-1, 1)$. 
\textcolor{black}{
Details of the 17 positions of the IMU sensors attached to the participants are described in \cite{huang2018deep}. For the reproducibility of the results in our paper, please refer to our GitHub repository at: \url{https://github.com/hieunq95/compressive-sensing-imu}.
}

By following the training process illustrated in Fig.~\ref{fig:vae-training}, the desired signal $\mathbf{x}^*$ is represented as $\mathbf{x}_t \in \mathbb{R}^{204}$. Hereafter, we remove the time step $t$ notation for the sake of simplicity. The signal $\mathbf{x}^*$ is then multiplied with the measurement matrix $\mathbf{A}$ to get the signal $\mathbf{y}$ with fewer features, i.e., $m < 204$. The signal $\mathbf{y}$ is then passed through a simulated channel with Gaussian noise $\boldsymbol{\eta} \in \mathbb{R}^{m}$. The noise vector $\boldsymbol{\eta}$ has $m$ elements, and each element follows a Gaussian distribution $\mathcal{N}(0, \sigma_N^2)$. 

The noisy signal $\mathbf{\hat{y}} \in \mathbb{R}^{m}$ is then used as the input of the VAE. The reconstructed signal at the output of the VAE is $\mathbf{\hat{x}} \in \mathbb{R}^{204}$. For this, we design the network architecture of the VAE as follows. The encoder is a fully connected network with an input layer having $m$ neurons, one hidden layer having 64 neurons, and one latent layer having 10 neurons. The decoder is a fully connected network that has two hidden layers, each of which has 64 neurons. 
\textcolor{black}{Finally, the output layer has 204 neurons, which are
equivalent to the number of features generated from the original IMU signals}. The activation function used for the hidden layers is ReLu and the activation function used for the output layer is Tanh. 
\textcolor{black}{
The ReLu activation function is selected due to its simplicity, computational efficiency, and ability to mitigate the vanishing gradient problem. Tanh is applied for the output layer as the reconstructed signal is being bounded within (-1, 1). Based on our experimental findings, we have determined that training the model for 50 epochs with a batch size of 60 leads to consistent and stable outcomes.}
We train the model for 50 epochs with a batch size of 60. We then use the trained model to evaluate the performance on the test set with the same batch size. All the parameter settings are described in Table \ref{tab:parameters}.

\subsubsection{3D Avatar Model}
\label{subsubsec:smpl}
Based on the reconstructed signals from the proposed CS-VAE model, we further transform the signals into a 3D human avatar. For this, we use the non-commercial Skinned Multi-Person Linear Model (SMPL) model in \cite{loper2023smpl}. SMPL is a parametrized model of a 3D human body template that takes 72 pose parameters and 10 shape parameters, denoted by $\mathbf{p} \in \mathbb{R}^{72}$ and $\mathbf{s} \in \mathbb{R}^{10}$, respectively, and returns a mesh with 6,890 vertices in a 3D space. By adjusting the pose and shape parameters, i.e., $\mathbf{p}$ and $\mathbf{s}$, we can animate the 3D avatars that mimic the physical shapes and movements of human users. Further details of the SMPL model can be found in \cite{loper2023smpl}. 

To transform the IMU signals $\mathbf{\hat{x}} \in \mathbb{R}^{204}$ into a pose parameter $\mathbf{p} \in \mathbb{R}^{72}$, we use another VAE to learn the mapping function $F(\mathbf{\hat{x}}):\mathbb{R}^{204} \rightarrow \mathbb{R}^{72}$. This mapping function helps us to transform IMU signals into the input of the SMPL model \cite{huang2018deep}. For example, the changes of acceleration and orientation of the IMU data (i.e., $\mathbf{\hat{x}}$ varies) from the left wrist and left elbow of the user will make the 3D avatar move its left arm (i.e., $\mathbf{p}$ varies). In other words, given the reconstructed signals, we can create a human avatar in a virtual 3D space with a specific pose. Note that we keep the shape parameters $\mathbf{s}$ of the SMPL model as constant numbers for the sake of simplicity as body shape modeling is not our focus.
In the following, we first present the performance evaluation of the signal reconstruction with our proposed framework. Secondly, we show that our reconstructed signals are more robust in creating 3D human avatar poses. Lastly, we show a simple but effective method to make smooth animated motions for the avatars without using input signals, thanks to the ability of the generative model. 

\subsubsection{Baseline Approaches}

\begin{table}[t]
\begin{tabular}{c|l|l}
\hline 
\textbf{Notation}		&\textbf{IMU Parameters} 			& \textbf{Values}\\
\hline
$n$ 	& Dimension of signals 				& $204$ features			\\ 
\hline
$m$ 	& Number of linear						& $[48:192]$ 				\\
		& measurements							& features					\\
\hline
		& Frame per seconds						& 60 fps \cite{myn2015xsens}			\\
\hline
$P_T$	& Transmit power					& $0.1$ Watt 					\\
			&											& 	\cite{myn2015xsens, cc2591}	             \\
\hline
$\sigma_N$	& 		Standard deviation of noise										& $[1: 500]$			\\
	&												& $\times 10^ {-4}$ \\
\hline
$d$	&		Parameter	of $\mathbf{A}$				& 2						\\  
	    &		  in Proposition \ref{prop:main} &					\\
\hline
		& \textbf{Algorithmic Parameters}		& \textbf{Values}\\
\hline 
		& Learning rate  		& 			$10^{-4}$							\\
		&	(Adam optimizer)   &												\\
\hline
		& KL divergence weight					& $10^{-5}$					\\
\hline
$\lambda$ & $l_1$ penalty						& $10^{-5}$					\\
\hline
		& Number of training epochs			& 50						\\
\hline
\end{tabular}
\caption{Parameter settings.}
\label{tab:parameters}
\end{table}

We evaluate the performance of our proposed framework, denoted by CS-VAE, and other baseline approaches. The considered baseline approaches are (i) Lasso (Lease absolute shrinkage and selection operator) \cite{tibshirani1996regression} and (ii) DIP (Deep Inertial Poser) \cite{huang2018deep}. 

\textbf{Lasso} is a widely adopted algorithm for solving $l_1$ penalized optimization problem in compressive sensing \cite{foucart2013invitation, choi2017compressed}. 
It is a regression analysis technique that incorporates an $l_1$ penalty term into the optimization problem. In particular, the solution of Lasso is obtained by solving the optimization problem in (\ref{eq:lasso-min}), which is equivalent to solving the Lagrangian in (\ref{eq:lasso-min-lambda}). As observed from equation (\ref{eq:lasso-min-lambda}), Lasso solves an optimization problem that involves the $l_1$ penalty term $\lambda \|\mathbf{x}\|_1$, which is similar to the term $\lambda \|G(\mathbf{z})\|_1$ in the loss function $L(\mathbf{z})$ in the proposed Algorithm \ref{algo:cs-vae}. For a fair comparison, we use the same value for the $l_1$ penalty $\lambda=10^{-5}$ for both Lasso and CS-VAE approaches. 

Note that by using the measurement matrix $\mathbf{A}$ that follows the Proposition \ref{prop:main}, the setting of Lasso in (\ref{eq:lasso-min-lambda}) is now constrained with transmit power in equation (\ref{eq:main-optimization}b). In our later experiments, we empirically show that the power constraint in (\ref{eq:main-optimization}b) makes the optimization problem much more challenging, resulting in Lasso's failures in reconstructing the original signals. For a more comprehensive evaluation, we also introduce a relaxed version of Lasso, denoted by the notation ``Lasso w.o.$P_T$" (Lasso without power constraint $P_T$). In this relaxed version of Lasso, we remove the power constraint in (\ref{eq:main-optimization}b) and use Lasso to recover the original signals. This can be done by replacing $\mathbf{A}$ in (\ref{eq:lasso-min-lambda}) with another unconstrained measurement matrix $\mathbf{B}$ with Gaussian entries. We use a similar measurement matrix as in \cite{bora2017compressed, dhar2018modeling} in which elements of $\mathbf{B} \in \mathbb{R}^{m \times n}$ are $B_{ij} \sim \mathcal{N}(0, 1/m)$.
Without the power constraint, ``Lasso w.o.$P_T$" is expected to produce the optimal results as upper bounds for our evaluation. As shown later, our proposed CS-VAE approach achieves comparable results to the optimal solutions obtained by ``Lasso w.o.$P_T$". Notably, the competitive results of our approach also come with an order of magnitude faster than Lasso in terms of decoding time, i.e., the time to find the solution for the optimization problem given a batch of input data.

\textbf{DIP} is a deep learning approach for reconstructing body pose from a fixed set of IMU sensors. The main idea of DIP and other approaches in this line of work, e.g., \cite{huang2018deep, yi2022physical}, is to reconstruct the full body pose, e.g., using SMPL model, from measurements of IMU sensors placed on the important joints of the human body, i.e., head, spine, two wrists, and two knees. The common approaches of such frameworks are using recurrent neural networks to access the entire training set during the training time and using a shorter time window at the test time. The number of IMU sensors can be reduced to 3 sensors, e.g., sensors placed on the head and two wrists, but such an approach needs an extensive motion database and physics-based simulation engine \cite{winkler2022questsim}.

In comparison with our proposed CS-VAE framework, the aforementioned works in \cite{huang2018deep, yi2022physical, winkler2022questsim} can be viewed as an underdetermined system in which the deep learning models try to recover the full body pose given signals from a few IMU sensors. As such, instead of using a matrix multiplication for downsampling the data, we can impose the training process of DIP by manually selecting the IMU sensors from the 17 IMU sensors in the dataset. 
Note that in the following results, we use the same VAE's architecture and training loss in Fig.~\ref{fig:vae-training} for the DIP baseline approach, rather than using the recurrent neural network in \cite{huang2018deep}. The main reason is that the recurrent neural network needs to see the entire training set during the training process, which may yield advantages and inappropriate comparisons with CS-VAE and Lasso. With similar architecture and training with batches of signals, it is a more reasonable comparison for DIP against CS-VAE and Lasso. The details of simulation parameters are described in Table \ref{tab:parameters}.

\subsection{Simulation Results}
\subsubsection{Impacts of the number of measurements}
\begin{figure}[t]
\centering
\includegraphics[width=0.8\linewidth]{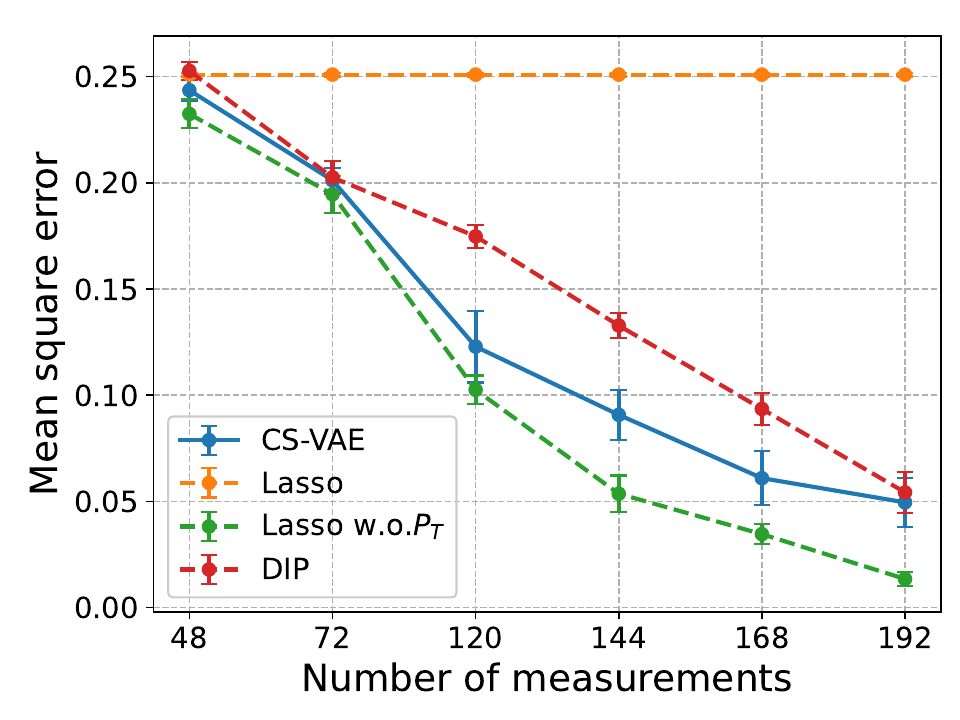}
\caption{Mean square error of reconstructed signals when the number of measurements $m$ increases.}
\label{fig:m-varies}
\end{figure}

We evaluate the performance of the proposed framework when the number of measurements $m$ increases from 48 measurements to 192 measurements, which are approximately $23\%$ and $94\%$ of the total number of the IMU orientation and acceleration features, respectively. We select the number of measurements in Fig.~\ref{fig:m-varies} to be divisible by 12. The only reason for this selection is that it is easier for the DIP baseline framework as the DIP framework needs to work with the set of IMU sensors in which each IMU has 12 features of measurements (9 features for the orientation and 3 features for the acceleration). Unlike DIP, the CS-VAE and Lasso approaches are flexible with any arbitrary number of measurements. The error bars in Figs. \ref{fig:m-varies}, \ref{fig:noise}, and \ref{fig:batch-varies} are equivalent to half of the standard deviations from the mean values.

As observed from Fig.~\ref{fig:m-varies}, the Mean Square Error (MSE) values in most approaches decrease when the number of measurements increases, except for Lasso as it fails to recover the signals. This observation about Lasso shows that the power constraint makes the setting much more difficult to obtain the results. When we remove the power constraint, the relaxed optimization problem can be effectively solved with Lasso, as illustrated by the Lasso w.o.$P_T$ baseline. Recall that we use Lasso w.o.$P_T$ as an upper bound for comparison. The results show that our CS-VAE approach achieves the highest performance, i.e., low MSE values, which is closest to the performance of the upper bound solution, i.e., Lasso w.o.$P_T$. 

The reason for the inferior performance of DIP to CS-VAE can be explained as follows. The linear projection of DIP from 204 measurements into a lower number of measurements only preserves the completeness of orientation and acceleration features. For example, the linear projection of DIP with $m=120$ measurements makes it equivalent to the data from 10 IMU sensors in which each IMU preserves the full 12 features of orientation and acceleration. With compressive sensing technique in other approaches, the completeness of such 12 features no longer holds as the linear projection is performed by the measurement matrix $\mathbf{A}$, which may preserve the features with the sparse values rather than keeping the full orientation and acceleration values of certain IMU sensors. We observe that our newly proposed design of the measurement matrix $\mathbf{A}$ is the key factor making the CS-VAE approach work well with noisy and sparse signals, given a simple neural network's architecture of VAE.

\subsubsection{Impacts of the channel noise}
\begin{figure}[t]
\centering
\includegraphics[width=0.8\linewidth]{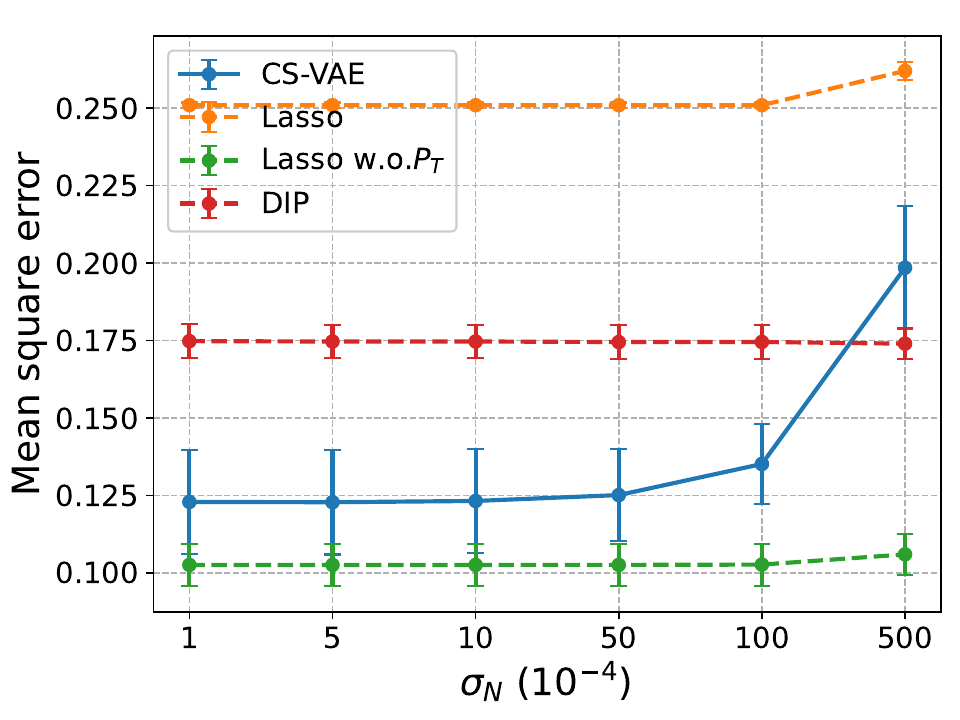}
\caption{Mean square error of reconstructed signals when channel noise power increases.}
\label{fig:noise}
\end{figure}

Next, we evaluate the performance of the approaches under different channel noise power values. As we consider the Gaussian channel, the noise's power is equivalent to the variance of the Gaussian noise, i.e., $\sigma_N^2$ in (\ref{eq:hat-y}) \cite{cover2006elements}. We fix the power $P_T$ and increase the standard deviation of the noise from $10^{-4}$ to $500 \times 10^{-4}$ to obtain the results in Fig.~\ref{fig:noise}. As observed from Fig.~\ref{fig:noise}, the proposed CS-VAE approach achieves better performance under most of the considered scenarios, compared to the Lasso and DIP approaches. Similar to the observation in the previous setting, Lasso fails to reconstruct the signals regardless of the noise level. The relaxed version of Lasso, i.e., Lasso w.o.$P_T$ achieves the highest performance as it is not constrained by the power transmission. 

We observe that under high noise's power value, i.e., $\sigma_N = 500 \times 10^{-4}$, CS-VAE performs worse than the DIP approach. The reason is that the design of matrix $\mathbf{A}$ can bound the signal $\mathbf{y} = \mathbf{Ax}$ with ${\frac{1}{m} \|\mathbf{y}\|_2^2 \leq P_T}$, but this also restricts the signal $\mathbf{y}$ into suboptimal power region. With DIP, we adopt the power normalization strategy in \cite{bourtsoulatze2019deep} in which the value of $\mathbf{y}$ is normalized by its $l_2$ norm $\|\mathbf{y}\|_2$. As the results suggest, this power normalization can be effective with high noise levels but it becomes less robust with low noise values. This is also the reason we design a new measurement matrix $\mathbf{A}$ rather than following this power normalization scheme as it yields a nonlinear projection from $\mathbf{x}$ to $\mathbf{y}$, which is in contrast to the linear projection idea in compressive sensing. Nevertheless, the CS-VAE achieves more robust and better performance than Lasso and DIP, and the results are closest to Lasso w.o.$P_T$.

\subsubsection{Decoding latency with respect to the number of input samples}

\begin{figure}[t]
\centering
\includegraphics[width=0.8\linewidth]{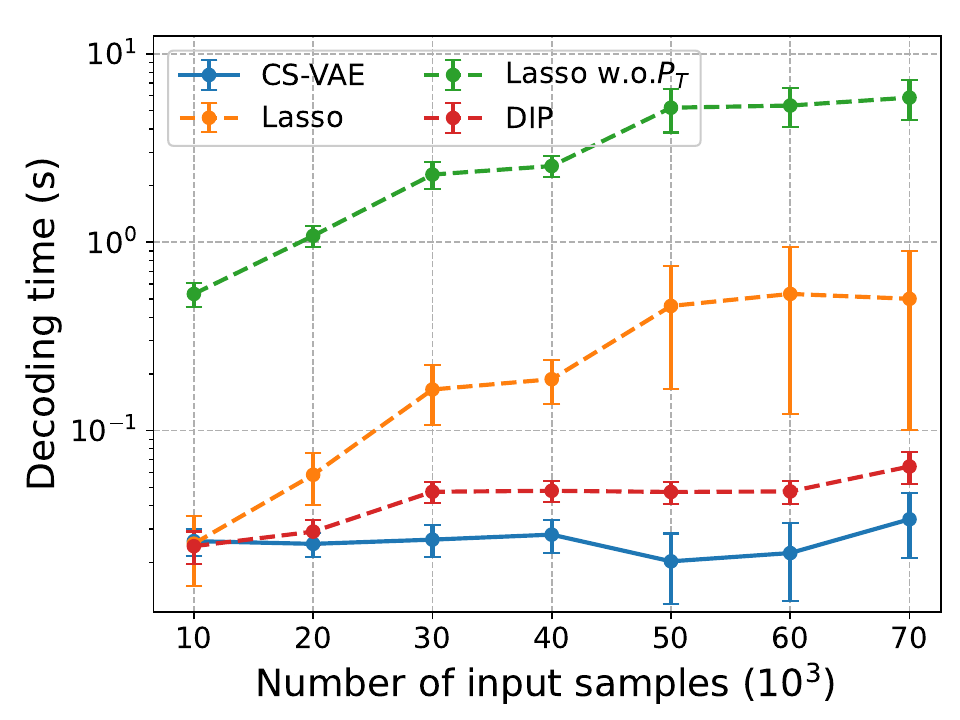} 
\caption{Decoding time at the receiver when the number of input samples increases.}
\label{fig:batch-varies}
\end{figure}

Next, we investigate the decoding time at the receiver under different sizes of the input samples in Fig.~\ref{fig:batch-varies}. The number of input samples $\mathbf{\hat{y}}$ fed into the VAE is equivalent to the number of measurements $m$ multiplied by the batch size. Note that in Fig.~\ref{fig:batch-varies}, we use the approximate values for the number of input samples in the x-axis for ease of illustration. The exact values are calculated by $m \times b$, where $m$ is the number of measurements and $b$ is the batch size. For example, with $10 \times 10^3$ input samples in Fig.~\ref{fig:batch-varies}, we use $m=168$ and $b=60$ to produce $10,080$ input samples.  These $10,080$ input samples are actually $14 \times 12 \times 60$ samples, where $14$ is the number of IMU sensors, $12$ is the number of features generated by each IMU sensor per frame, and $60$ is the number of frames per second. In other words,  a sequence of $10 \times 10^3$ input samples is equivalent to data generated by the IMU sensors in one second. Similarly, a sequence of $70 \times 10^3$ input samples is equivalent to data generated by the IMU sensors in 7 seconds.

The transmission of a batch of data samples is equivalent to the case when we collect a batch of samples after a period of time and then transmit them to the receiver. For Lasso and Lasso w.o.$P_T$, this batch of samples can be effectively used as the input of the optimization solver as the Lasso model can handle matrix-like input data.  The decoding time is measured as the single forward operation from the input to the output of the pre-trained VAE models to obtain the reconstructed signals. With Lasso, the decoding time is measured as the time to find the solution for the optimization problem. We use the same central processing unit (CPU) to compute the decoding time for all approaches. 

As observed from Fig.~\ref{fig:batch-varies}, the decoding time values of the CS-VAE and DIP approaches are significantly lower than that of the Lasso approaches. The main reason is that the search space of Lasso's solver increases as the number of input samples increases. In contrast, the decoding time values of CS-VAE and DIP slowly increase with the number of input samples because the pre-trained VAE models only need to forward the received signals through the deep neural network layers to obtain the reconstructed signals. Notably, the cost of finding high accuracy for reconstructed signals of Lasso w.o.$P_T$ makes it significantly slower than the deep learning approaches.

It is worth mentioning that the CS-VAE and DIP approaches need to be pre-trained before being applied to get the results in Fig.~\ref{fig:batch-varies} while the Lasso approach does not have this pre-training process. However, the pre-training process can be greatly facilitated through modern graphics processing unit (GPU)  training. The observation in this experiment also suggests the real-time decoding capability of the CS-VAE approach with lower decoding latency. For example, with a number of input samples of $10^4$, which is equivalent to a sequence data of one second, the decoding time of the CS-VAE approach is approximately $8 \times 10^{-2}$ second. We observe that although sharing the same network architecture with the CS-VAE approach, the DIP approach experiences slightly higher decoding time. The main reason is that the implementation of the linear projection matrix of DIP requires a few extra matrix transformation steps and de-transformation steps. However, this difference in the decoding time is not significant.

\subsubsection{3D pose estimation from reconstructed signals}

\begin{figure}[t]
\centering
\includegraphics[width=1.0\linewidth]{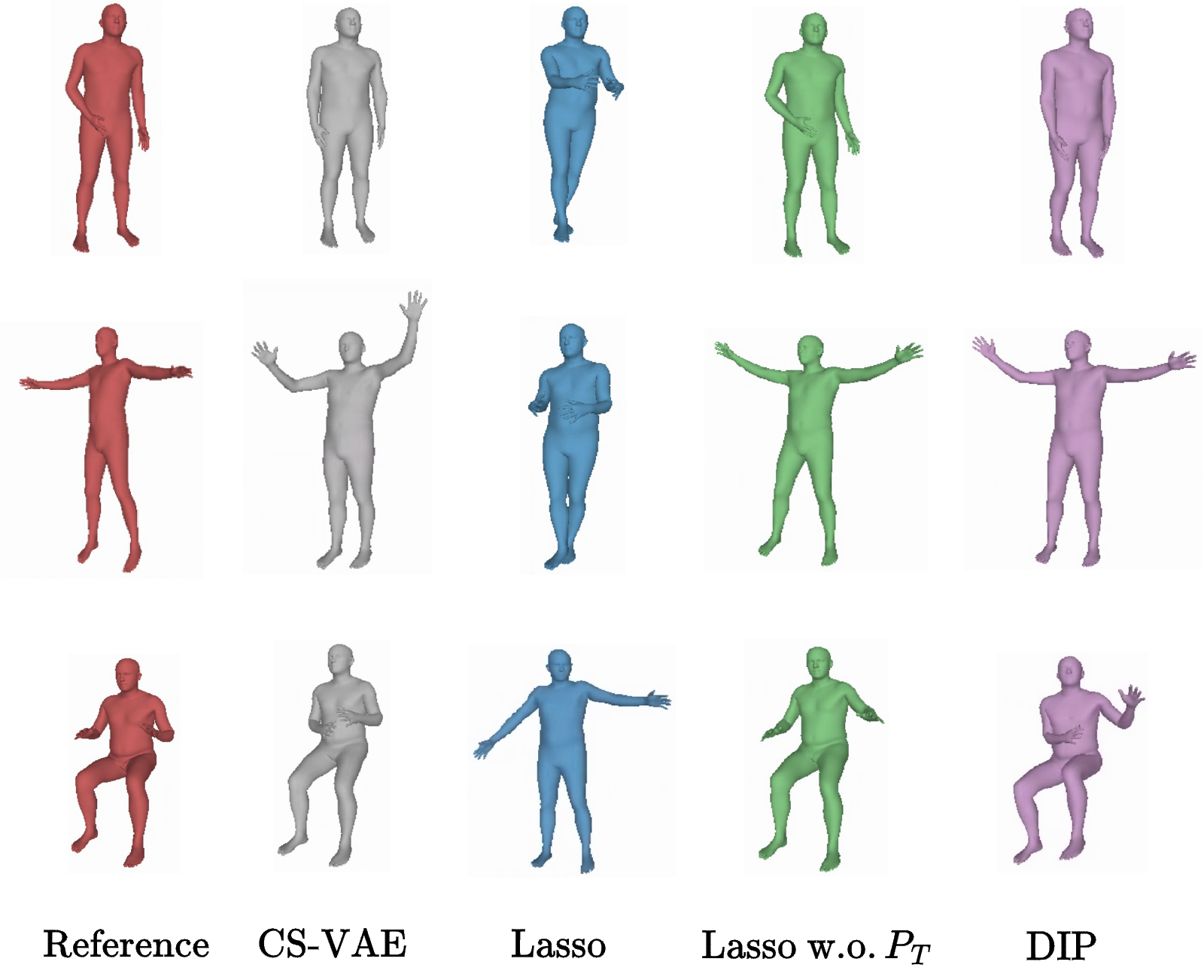}
\caption{Reconstructed 3D poses from noisy and compressed $m=168$ measurements and noise $\sigma_N = 10 \times 10^{-4}$.}
\label{fig:recons-pose}
\end{figure}

In Fig.~\ref{fig:recons-pose}, we draw random poses from the reconstructed signals in the test set. The poses are generated from the SMPL model and the pre-trained VAE model (modeled as a function $F(\mathbf{\hat{x}}): \mathbb{R}^{204} \rightarrow \mathbb{R}^{72}$) as described in Section \ref{subsubsec:smpl}. In this experiment, we reconstruct the poses from 168 measurements (i.e., 82\% of the total signals) and the channel noise power $\sigma_N = 10 \times 10^{-4}$. As observed from Fig.~\ref{fig:recons-pose}, our reconstructed poses are more accurate compared with the Lasso and DIP approaches. As Lasso w.o.$P_T$ is considered as the upper bound of all the signal reconstruction scenarios, Fig.~\ref{fig:recons-pose} clearly shows that the reconstructed 3D poses by Lasso w.o.$P_T$ (green poses in the fourth column) achieve the most similarity to the ground truth reference poses (red poses in the first column). The 3D poses obtained by CS-VAE can accurately mimic the body poses of the references with slight errors in arm movements, e.g., the pose in the second row. With Lasso, as suggested by the poor signal reconstruction performance, the 3D poses obtained by Lasso fail to mimic the reference poses. In our experiment, the received signals are down-sampled and contain additive noise and the models do not have access to the full training set as in \cite{huang2018deep}, resulting in poor performance of the DIP approach. 

\subsubsection{Pose interpolation without input signals}
\begin{figure}[t]
\centering
\includegraphics[width=1.0\linewidth]{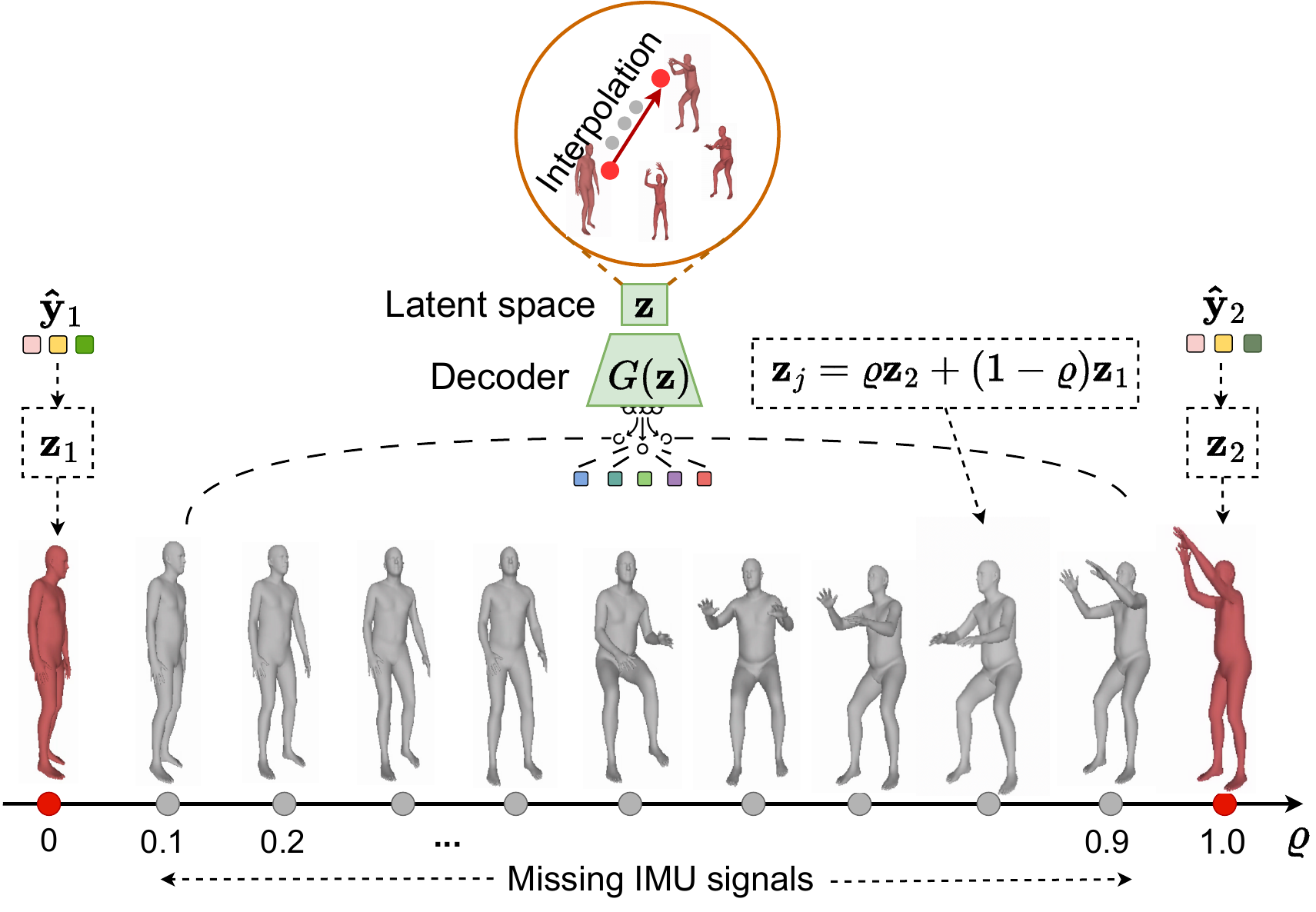}
\caption{Interpolation between key poses (red avatars) without using input IMU signals. The gray avatars represent the poses generated by the VAE when the input IMU signals are missing, e.g., due to transmission loss.}
\label{fig:interpolation}
\end{figure}

We further use the pre-trained latent space and the decoder of the VAE to generate novel synthesis poses to animate the avatar. The ability to generate synthesis data is one of the most important features of generative models, which has not been well demonstrated in the literature of wireless systems. 
In particular, we consider a simple pose interpolation task as follows. Given two key poses, illustrated by the left pose and right pose in red color in Fig.~\ref{fig:interpolation}, we aim to create a smooth transition between these two key poses by generating the immediate light gray colored poses. The need for creating the intermediate poses is important as the IMU signals might be lost during the transmission over the severe lossy wireless channels. In such a case, if the receiver cannot fill the missing poses or the transmitter does not retransmit the data, the user may experience motion sickness in the virtual 3D environment, thus decreasing the quality of experience. 

Given the above setting, as illustrated in Fig.~\ref{fig:interpolation}, we obtain the two key poses by reconstructing the IMU signals similar to the previous experiments. Let's denote the received signals $\mathbf{\hat{y}}_1$ and $\mathbf{\hat{y}}_2$ corresponding to the two key poses, respectively. The VAE's encoder can map the signals into two corresponding vectors in the latent space that are $\mathbf{z}_1 = q_{\boldsymbol{\phi}}(\mathbf{z} | \mathbf{y} = \mathbf{\hat{y}}_1)$ and $\mathbf{z}_2 = q_{\boldsymbol{\phi}}(\mathbf{z} | \mathbf{y} = \mathbf{\hat{y}}_2)$. As a result, the reconstructed signals corresponding to the two key poses are $\mathbf{\hat{x}}_1 = p_{\boldsymbol{\theta}}(\mathbf{x | z} = \mathbf{z}_1)$ and $\mathbf{\hat{x}}_2 = p_{\boldsymbol{\theta}}(\mathbf{x | z} = \mathbf{z}_2)$. To simply interpolate between the two latent vectors $\mathbf{z}_1$ and $\mathbf{z}_2$, we use the intermediate latent vectors $\mathbf{z}_j = \varrho \mathbf{z}_2 + (1 - \varrho) \mathbf{z}_1$ where $\varrho \in [0, 1]$ is the interpolation parameter. Intuitively, $\varrho = 1$ makes the intermediate latent vector $\mathbf{z}_j = \mathbf{z}_2$, resulting in the pose similar to the key pose on the right of the Fig.~\ref{fig:interpolation}. Similarly, $\varrho = 0$ imposes the vector $\mathbf{z}_j = \mathbf{z}_1$. The arbitrary value of $\varrho \in [0, 1]$ creates a latent vector that is a linear combination of the two vectors $\mathbf{z}_1$ and $\mathbf{z}_2$. As a result, we can make a smooth transition of the avatar poses by simply increasing the value of $\varrho$, as shown in Fig.~\ref{fig:interpolation}. As described, there is no need for the input signals when interpolation takes place in the learned latent space $\mathbf{z}$. 
Similar linear interpolation techniques have been explored to create synthesized data in the various domains \cite{berthelot2018understanding}. Our experiment shows the potential extension of the proposed framework to future VR/XR applications in conjunction with generative modeling where the synthesized data can be utilized. 

\section{Conclusion}
\label{sec:conclusion}
In this paper, we have developed a novel framework for 3D human pose estimation from IMU sensors with generative model-based compressive sensing. The proposed framework helps the IMU sensors reduce the amount of information exchanged with the receiver, thus further enhancing channel utilization and encoding-decoding efficiency. At the receiver's side, we have employed a deep generative model, i.e., a VAE, that can recover the original signals from noisy compressed samples. With the ability of the generative model at the receiver, we have achieved an order of magnitude faster than Lasso in terms of decoding latency.
We have further demonstrated that the proposed framework can learn a latent representation space and generate synthetic data samples, making it possible to fulfill missing data features (e.g., due to lossy transmissions) without using input data from the IMU sensors. Interesting findings suggest that the proposed generative model-based compressive sensing framework can achieve state-of-the-art performance in challenging scenarios with severe noise and fewer number of measurements, compared with other optimization and deep learning approaches. 
\textcolor{black}{
One potential research direction from this work is extending the designed measurement matrix for other complicated channel models, such as additive noise fading channel, Rayleigh fading channel, and Rician fading channel. Multiple-access channels with interference can also be further considered with additional constraints on the design of the measurement matrix.
}

\appendices 
\section{Proof of Proposition \ref{prop:main}}
\label{appendix}
The purpose of the Proposition \ref{prop:main} is to construct a measurement matrix that satisfies the S-REC property in Definition \ref{def:s-rec} and the power constraint in (\ref{eq:main-optimization}) so that the recovered signal of the generative model is unique and accurate with high probability. In the following, we prove that if each element $A_{ij}$ (element $j$-th of the $i$-th row of matrix $\mathbf{A}$) follows a normal distribution $A_{ij} \sim \mathcal{N}\Big(0, \frac{P_T}{n^2 d^2 (d \sigma_x + \mu_x)^2}\Big)$, the matrix $\mathbf{A} \in \mathbb{R}^{m \times n}$ satisfies the S-REC property and power constraint. The first part of this section proves the guarantee of the power constraint. The second part proves the S-REC property of the measurement matrix.

\subsection{Proof of $\mathbf{y} = \mathbf{Ax}$ guaranteeing the power constraint of a Gaussian channel}
First, we need prove that the matrix $\mathbf{A}$ satisfies that power constraint
\begin{equation}
\frac{1}{m} \|\mathbf{y}\|_2^2 \leq P_T.
\end{equation}
By using $l_2$ norm definition in (\ref{eq:lp-definition}), the power constraint can be rewritten as \cite[Equation 9.2, Chapter 9]{cover2006elements}:
\begin{equation}
\frac{1}{m} \sum_{i=1}^m |y_i|^2 \leq P_T.
\end{equation}
Recall that we have a $n$-dimensional vector $\mathbf{x} = [x_1, x_2, \ldots, x_j, \ldots, x_n]^{\top}$ and the $m$-dimensional vector $\mathbf{y} = [y_1, y_2, \ldots, y_i, \ldots, y_m]^{\top}$, where the superscript $\top$ denotes the transpose. The measurement matrix $\mathbf{A} \in \mathbb{R}^{m \times n}$ is defined by:
\begin{equation*}
\mathbf{A}=\left[\begin{array}{cccc}
A_{11} & A_{12} & \cdots & A_{1 n} \\
A_{21} & A_{22} & \cdots & A_{2 n} \\
\vdots & \vdots & \ddots & \vdots \\
A_{m 1} & A_{m 2} & \cdots & A_{m n}
\end{array}\right].
\end{equation*}

Using matrix calculation, we have $y_i = \sum_{j=1}^n A_{ij} x_j$ with $i = 1, 2, \ldots, m$. Following the definition of $l_p$ norm in (\ref{eq:lp-definition}), we have 
\begin{subequations}
\label{eq:cauchy-apply}
\begin{align}
\frac{1}{m} \|\mathbf{y}\|_2^2 & = \frac{1}{m} \sum_{i=1}^m |y_i|^2 \\
 & = \frac{1}{m} \sum_{i=1}^m \Big(\sum_{j=1}^n A_{ij}x_j \Big)^2 \\
 & \leq \frac{1}{m} \sum_{i=1}^m \Big[\big(\sum_{j=1}^n A_{ij}^2\big) \big(\sum_{j=1}^n x_j^2\big) \Big],
\end{align}
\end{subequations}
where (\ref{eq:cauchy-apply}c) directly applies Cauchy–Schwarz inequality, i.e., $\Big(\sum_{j=1}^n A_{ij}x_j \Big)^2 \leq  \big(\sum_{j=1}^n A_{ij}^2\big) \big(\sum_{j=1}^n x_j^2 \big)$. 
In the following, we derive the bounds for the two terms inside (\ref{eq:cauchy-apply}c) which are $\sum_{j=1}^n A_{ij}^2$ and $\sum_{j=1}^n x_j^2$. For this purpose, we use Chebyshev's inequality \cite[Chapter 3, Equation 3.32]{cover2006elements}, which can be stated as follows.

\begin{definition}[Chebyshev's inequality]
Let $X$ be a random variable with mean $\mu$ and variance $\sigma^2$. For any $\varepsilon > 0$, 
\begin{equation*}
\mathbb{P}\big(|X - \mu| > \varepsilon\big) \leq \frac{\sigma^2}{\varepsilon^2}.
\end{equation*}
\end{definition}

In our setting, we are more interested in the central limits, i.e., the distances away from the mean values, of the random variables, i.e., $x_j$ and $A_{ij}$. Let's $\varepsilon = d \sigma$ for real number $d > 0$, Chebyshev's inequality can be rewritten as
\begin{equation}
\mathbb{P}\big(|X - \mu| > d \sigma\big) \leq \frac{1}{d^2}.
\label{eq:chebyshev-d}
\end{equation}

Equivalently, we can bound the absolute value of $|X - \mu| \leq d \sigma$ with probability is at least $1 - \frac{1}{d^2}$, i.e., $\mathbb{P}\big(|X - \mu| \leq d \sigma\big) > 1 - \frac{1}{d^2}$. By choosing the value of $d$, we can bound the value of $X$ within a certain distance away from its mean with known probability. 
For sufficiently large $d$, we have the following inequalities
\begin{equation}
\mu - d \sigma \leq X \leq \mu + d \sigma,
\label{eq:central-limits}
\end{equation}
with high probability.

Now, by using inequalities in (\ref{eq:central-limits}) for random variables $x_j$ associated with mean $\mu_x$ and variance $\sigma_x^2$, and $A_{ij}$ associated with zero-mean and variance $\sigma_a^2$, we have the following inequalities
\begin{subequations}
\begin{align}
& \quad -d \sigma_a \leq A_{ij} \leq d \sigma_a, \\
& \mu_x - d \sigma_x \leq x_j \leq \mu_x + d \sigma_x,
\end{align}
\end{subequations}
with high probabilities.
By taking the square of $A_{ij}$ and $x_j$, we have
\begin{subequations}
\begin{align}
\quad & A_{ij}^2 \leq d \sigma_a^2, \\
\quad & x_j^2 \leq \max\big[(d \sigma_x^2 + \mu_x)^2, (- d \sigma_x + \mu_x)^2\big],
\end{align}
\end{subequations}
with high probabilities. Taking the sum over $n$ samples, we have 
\begin{equation}
\sum_{j=1}^n A_{ij}^2 \leq n d \sigma_a^2,
\label{eq:bound-sum-a}
\end{equation}
and 
\begin{equation}
\sum_{j=1}^n x_j^2 \leq \max\big[n (d \sigma_x^2 + \mu_x)^2, n (- d \sigma_x + \mu_x)^2\big],
\end{equation}
with high probabilities. As we empirically observe from the dataset that $\mu_x > 0$, the bound for the above equation can be simplified as 
\begin{equation}
\sum_{j=1}^n x_j^2 \leq n (d \sigma_x^2 + \mu_x)^2.
\label{eq:bound-sum-x}
\end{equation}
Replacing (\ref{eq:bound-sum-a}) and (\ref{eq:bound-sum-x}) in (\ref{eq:cauchy-apply}), we finally have
\begin{subequations}
\begin{align}
\frac{1}{m} \|\mathbf{y}\|_2^2 & \leq \frac{1}{m} \sum_{i=1}^m n d^2 \sigma_a^2 n (d \sigma_x + \mu_x)^2 \\
 & = n^2 d^2 (d \sigma_x + \mu_x)^2 \sigma_a^2. \\
\end{align}
\end{subequations}
As we want to have a power constraint $\frac{1}{m}\|\mathbf{y}\|_2^2 \leq P_T$, by choosing $n^2 d^2 (d \sigma_x + \mu_x)^2 \sigma_a^2 = P_T$, we have the variance of $A_{ij}$ is
\begin{equation}
\label{eq:final-sigma-a}
\sigma_a^2 = \frac{P_T}{n^2 d^2 (d \sigma_x  +\mu_x)^2}.
\end{equation}

The derivation above for $\sigma_a$ proves the Proposition \ref{prop:main}. In other words, by generating the measurement matrix $\mathbf{A}$ from the normal distribution with zero-mean and variance $\sigma_a^2$, the transmit power $\frac{1}{m} \|\mathbf{y}\|_2 \leq P_T$ can be achieved with high probability.

\subsection{Proof of $\mathbf{Ax}$ satisfying S-REC property}
Next, we prove that given the measurement matrix $\mathbf{A}$ that follows the Proposition \ref{prop:main} will satisfy the S-REC property in Definition \ref{def:s-rec}. In particular, the matrix $\mathbf{A}$ is said to satisfy the $\text{S-REC}(S_G, \gamma, \kappa)$, i.e., set-restricted eigenvalue condition of the set $S_G$ (defined in (\ref{eq:set-G})) with the parameters $\gamma > 0$ and $\kappa \geq 0$, if $\forall \mathbf{x}_1, \mathbf{x}_2 \in S_G$,
\begin{equation}
\|\mathbf{A}\big(G(\mathbf{z}_1) - G(\mathbf{z}_2)\big)\|_2 \geq \gamma \|G(\mathbf{z}_1) - G(\mathbf{z}_2)\|_2 - \kappa.
\label{eq:s-rec-agz}
\end{equation}

The range of the generator $G(\mathbf{z})$ can be easily bounded by the output layer of the deep neural network. In our experiments, we use the Tanh activation function as the output of the neural network. Therefore, we have a simple bound $-1 \leq G(\mathbf{z}) \leq 1$, which is similar to the bound of processed signal vectors $\mathbf{x}^*$.

Let's define a vector $\mathbf{v} = G(\mathbf{z}_1) - G(\mathbf{z}_2)$ with the bound of each element $v_j$ ($j=1, 2, \ldots, n$) of $\mathbf{v}$ is $-2 \leq v_j \leq 2$, (\ref{eq:s-rec-agz}) can be rewritten as 
\begin{equation}
\label{eq:s-rec-v}
\|\mathbf{Av}\|_2 \geq \gamma \|\mathbf{v}\|_2 - \kappa.
\end{equation}

By using the definition of $l_p$ norm in (\ref{eq:lp-definition}), we have
\begin{subequations}
\label{eq:gamma-v-inequality}
\begin{align}
\gamma \|\mathbf{v}\|_2 & = \gamma \sqrt{\sum_{j=1}^n v_j^2} \\
& \leq \gamma \sqrt{4n} \\
& = 2 \gamma \sqrt{n},
\end{align}
\end{subequations}
where the inequality in the second line is obtained by the bound $-2 \leq v_j \leq 2$. As a result, we have the following inequality for the term on the right-hand side of (\ref{eq:s-rec-agz}):
\begin{equation}
\label{eq:kappa-inequality}
\gamma \|\mathbf{v}\|_2 - \kappa \leq 2 \gamma \sqrt{n} - \kappa.
\end{equation}

To find a possible lower bound for $\|\mathbf{Av}\|_2$, we use the inequalities between the $l_1$ and $l_2$ norm, and then find the probabilistic lower bound of the $l_1$ norm based on Bernstein inequality. In particular, we apply the following inequality (see equations (A.3) and (A.4) in Definition A.2 of \cite{foucart2013invitation}):
\begin{equation}
\|\mathbf{Av}\|_2 \geq \frac{1}{\sqrt{m}} \|\mathbf{Av}\|_1.
\label{eq:l2-l1-inequality}
\end{equation}

Using the definition of $l_1$ norm in (\ref{eq:lp-definition}), (\ref{eq:l2-l1-inequality}) can be rewritten as
\begin{subequations}
\label{eq:generalized-triangle}
\begin{align}
\|\mathbf{Av}\|_2 & \geq \frac{1}{\sqrt{m}} \sum_{j=1}^n \big|A_{ij}v_j\big| \\
& \geq \frac{1}{\sqrt{m}} \Big|\sum_{j=1}^n A_{ij} v_j\Big|,
\end{align}
\end{subequations}
where the inequality in the second line is obtained by using the generalized triangle inequality. 
Next, we use a probabilistic lower bound for (\ref{eq:generalized-triangle}b) which applies Bernstein inequality (see Theorem 7.27 - Chapter 7 of \cite{foucart2013invitation}), i.e., given the measurement matrix $\mathbf{A}$, which has elements $A_{ij}$ are zero-mean sub-gaussian random variables, we have the following probabilistic lower bound of
\begin{equation}
\label{eq:bernstein-inequality}
\mathbb{P}\Big(\big|\sum_{j=1}^n A_{ij} v_j\big| \geq t\Big) \leq 2 \exp \Big({\frac{-t^2}{4c\|\mathbf{A}\|_2^2}}\Big),
\end{equation}
for $\forall t > 0$, where $c$ is a subgaussian parameter. Let's $t = t_0 \sqrt{m}$ with $\forall t_0 > 0$, (\ref{eq:bernstein-inequality}) can be rewritten as
\begin{subequations}
\label{eq:bernstein-after}
\begin{align}
& \mathbb{P}\Big(\big|\sum_{j=1}^n A_{ij} v_j\big| \geq \sqrt{m} t_0\Big) \leq 2 \exp \Big(\frac{-t_0^2 m^2}{4c\|\mathbf{A}\|_2^2}\Big) \\
\Rightarrow & \mathbb{P}\Big(\frac{1}{\sqrt{m}} \big|\sum_{j=1}^n A_{ij} v_j\big| \geq t_0\Big) \leq 2 \exp\Big(\frac{-t_0^2 m^2}{4c\|\mathbf{A}\|_2^2}\Big).
\end{align}
\end{subequations}
Using the inequality in (\ref{eq:generalized-triangle}), (\ref{eq:bernstein-after}b) becomes
\begin{equation}
\label{eq:av-final-bound}
\mathbb{P}\Big(\|\mathbf{Av}\|_2 \geq t_0\Big) \leq 2 \exp \Big(\frac{-t_0^2 m^2}{4c\|\mathbf{A}\|_2^2}\Big).
\end{equation}
By choosing $t_0 = 2 \gamma \sqrt{n} - \kappa$, (\ref{eq:s-rec-v}) can be written as
\begin{equation}
\|\mathbf{Av}\|_2 \geq 2 \gamma \sqrt{n} - \kappa,
\end{equation}
with probability $1 - 2 \exp \Big(\frac{(-2 \gamma m\sqrt{n} - m \kappa)^2}{4c\|\mathbf{A}\|_2^2}\Big)$. Applying inequality in (\ref{eq:kappa-inequality}), i.e., $2 \gamma \sqrt{n} - \kappa \geq \gamma \|\mathbf{v}\|_2 - \kappa$, we have
\begin{equation}
\|\mathbf{Av}\|_2 \geq \|\mathbf{v}\|_2 - \kappa
\end{equation}
with probability is at least $1 - 2 \exp \Big(\frac{(-2 \gamma m\sqrt{n} - m \kappa)^2}{4c\|\mathbf{A}\|_2^2}\Big)$. As $\mathbf{v}$ is defined by $\mathbf{v} = G(\mathbf{z}_1) - G(\mathbf{z}_2)$, finally, we have
\begin{equation}
\label{eq:s-rec-final-proof}
\|\mathbf{A}\big(G(\mathbf{z}_1) - G(\mathbf{z}_2)\big)\|_2 \geq \gamma \|G(\mathbf{z}_1) - G(\mathbf{z}_2)\|_2 - \kappa
\end{equation}
with probability is at least $1 - 2 \exp \Big(\frac{(-2 \gamma m\sqrt{n} - m \kappa)^2}{4c\|\mathbf{A}\|_2^2}\Big)$. This proves the S-REC($S_G, \gamma, \kappa$) property in (\ref{eq:s-rec-agz}). 

By combining (\ref{eq:s-rec-final-proof}) and (\ref{eq:final-sigma-a}), the proof is now completed.

\bibliography{refs}

\begin{IEEEbiography}[{\includegraphics[width=1in,height=1.25in,clip,keepaspectratio]{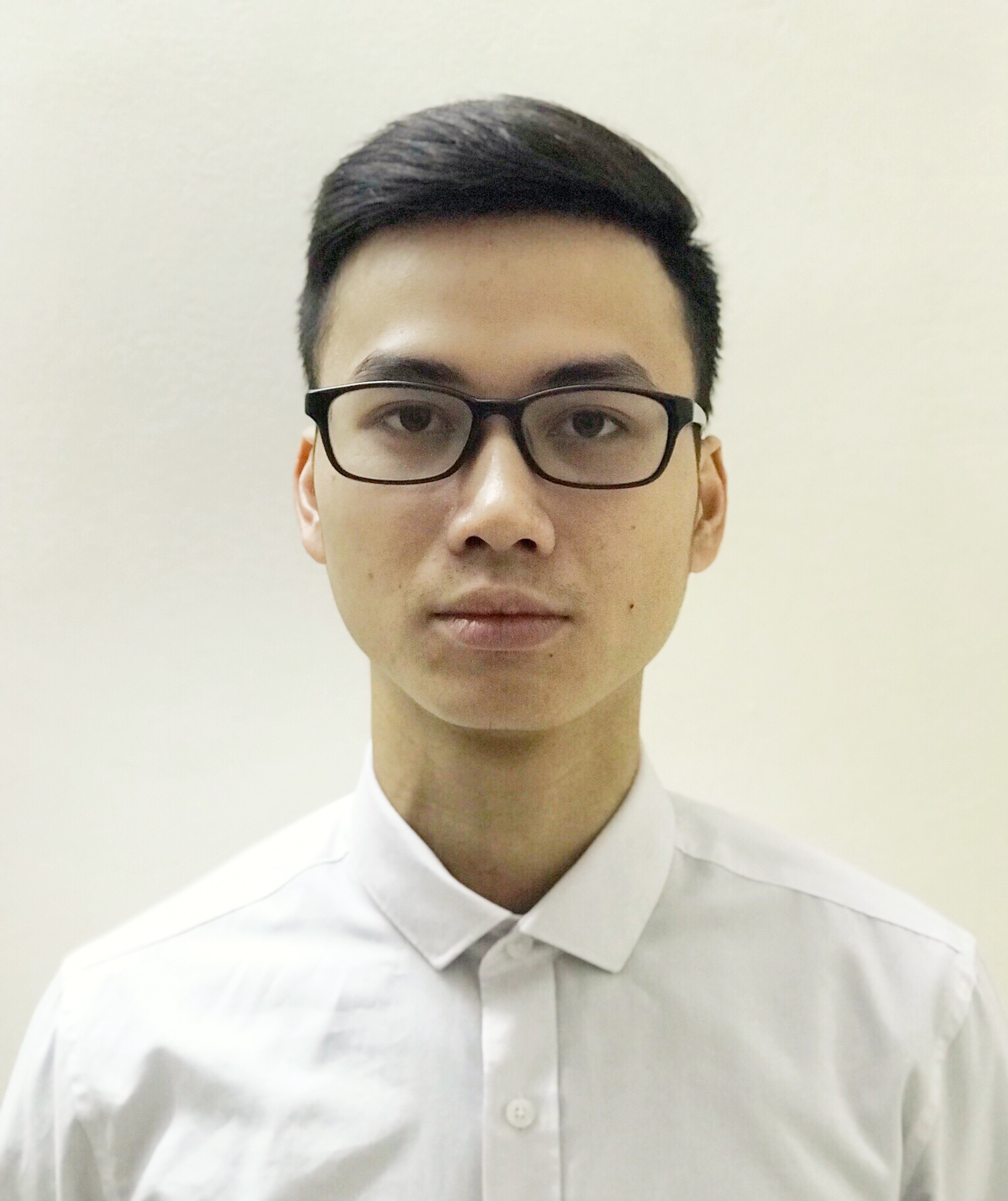}}]%
{Nguyen Quang Hieu} received the B.E. degree in Hanoi University of Science Technology, Vietnam in 2018. He is currently a Ph.D. student at School of Electrical and Data Engineering, University of Technology (UTS), Sydney, Australia. Before joining UTS, he was a research assistant at School of Computer Science and Engineering, Nanyang Technological University, Singapore. His research interest include wireless communications and machine learning.
\end{IEEEbiography}

\begin{IEEEbiography}[{\includegraphics[width=1in,height=1.25in,clip,keepaspectratio]{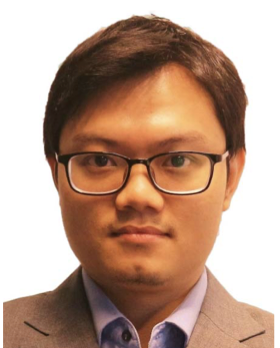}}]{Dinh Thai Hoang} (M'16, SM'22) is currently a faculty member at the School of Electrical and Data Engineering, University of Technology Sydney, Australia. He received his Ph.D. in Computer Science and Engineering from the Nanyang Technological University, Singapore 2016. His research interests include emerging wireless communications and networking topics, especially machine learning applications in networking, edge computing, and cybersecurity. He has received several precious awards, including the Australian Research Council Discovery Early Career Researcher Award, IEEE TCSC Award for Excellence in Scalable Computing for Contributions on “Intelligent Mobile Edge Computing Systems” (Early Career Researcher), IEEE Asia-Pacific Board (APB) Outstanding Paper Award 2022, and IEEE Communications Society Best Survey Paper Award 2023.  He is currently an Editor of IEEE TMC, IEEE TWC, IEEE TCCN, IEEE TVT, and IEEE COMST.\end{IEEEbiography}

\begin{IEEEbiography}[{\includegraphics[width=1in,height=1.25in,clip,keepaspectratio]{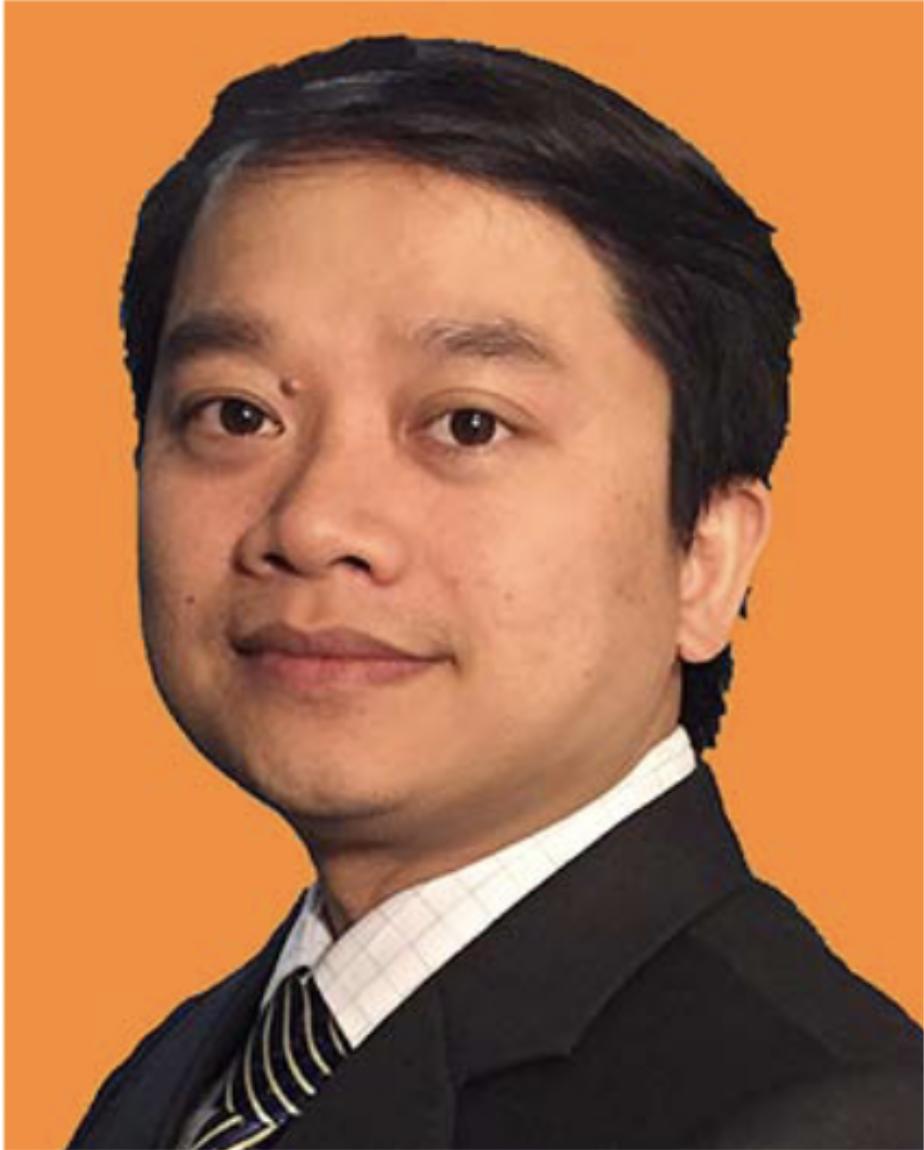}}]{Diep N. Nguyen}
 (Senior Member, IEEE) received the M.E. degree in electrical and computer engineering from the University of California at San Diego (UCSD), La Jolla, CA, USA, in 2008, and the Ph.D. degree in electrical and computer engineering from The University of Arizona (UA), Tucson, AZ, USA, in 2013. He is currently the Head of 5G/6G Wireless Communications and Networking Lab, Director of Agile Communications and Computing group, Faculty of Engineering and Information Technology, University of Technology Sydney (UTS), Sydney, NSW, Australia. Before joining UTS, he was a DECRA Research Fellow with Macquarie University, Macquarie Park, NSW, Australia, and a Member of the Technical Staff with Broadcom Corporation, CA, USA, and ARCON Corporation, Boston, MA, USA, and consulting the Federal Administration of Aviation, Washington, DC, USA, on turning detection of UAVs and aircraft, and the U.S. Air Force Research Laboratory, USA, on anti-jamming. His research interests include computer networking, wireless communications, and machine learning application, with emphasis on systems' performance and security/privacy. Dr. Nguyen received several awards from LG Electronics, UCSD, UA, the U.S. National Science Foundation, and the Australian Research Council. He has served on the Editorial Boards of the IEEE Transactions on Mobile Computing, IEEE Communications Surveys \& Tutorials (COMST), IEEE Open Journal of the Communications Society, and Scientific Reports (Nature's).\end{IEEEbiography}
 
\vspace{12cm}
\begin{IEEEbiography}
 [{\includegraphics[width=1in,height=1.25in,clip,keepaspectratio]{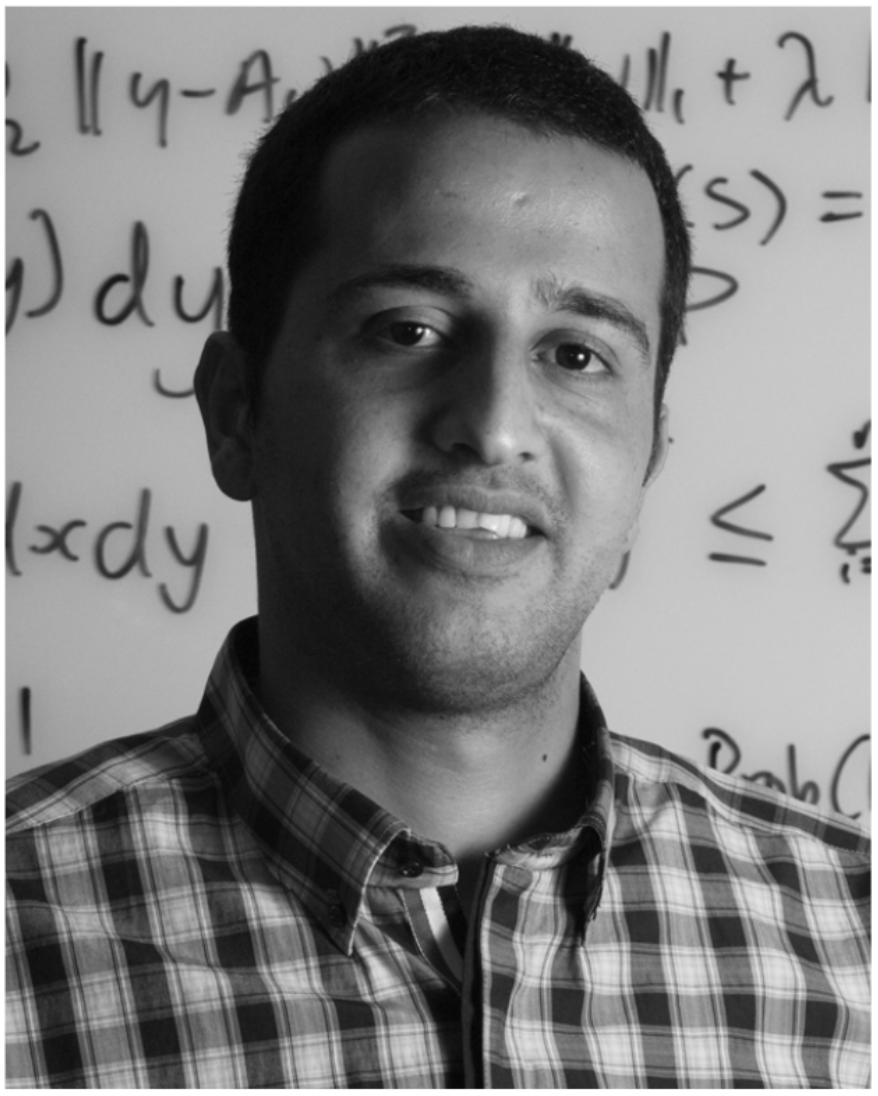}}]{Mohammad Abu Alsheikh}
(Senior Member, IEEE) received the B.Eng. degree in computer systems from Birzeit University, Palestine. He worked as a Software Engineer at a digital
advertising start-up and Cisco. Previously, he was a Postdoctoral Researcher with the Massachusetts Institute of Technology (MIT), USA. He is currently an Associate Professor and an ARC DECRA Fellow with the University of Canberra (UC), ACT, Australia. He designs and creates novel privacy-preserving Internet of Things systems that leverage both machine learning and convex optimization with applications in people-centric sensing, human activity recognition, and smart cities. His Ph.D. research at Nanyang Technological University (NTU), Singapore, focused on optimizing wireless sensor network's data collection.
\end{IEEEbiography}

\end{document}